\documentclass[prl,superscriptaddress, amsmath,amssymb, aps,twocolumn,longbibliography]{revtex4-1} 
\usepackage{graphicx}
\usepackage{dcolumn}
\usepackage{amsthm}
\PassOptionsToPackage{usenames,dvipsnames}{xcolor}
\usepackage{orcidlink}
\usepackage{bm}
\newcommand{\ket}[1]{|#1\rangle}
\newcommand{\bra}[1]{\langle#1|}

\pdfoutput=1
\usepackage{graphicx,graphics,epsfig,subfigure,times,bm,bbm,amssymb,amsmath,amsthm,mathrsfs,MnSymbol}
\usepackage{gensymb}
\usepackage{amsfonts}
\usepackage[matrix,frame,arrow]{xypic}
\usepackage{hyperref}
\hypersetup{
    pdfstartview=FitH,
    colorlinks=true,       
    linkcolor=red,          
    citecolor=magenta,        
    filecolor=magenta,      
    urlcolor=cyan,           
    runcolor=cyan}
\usepackage{epstopdf}
\usepackage{color}
\usepackage{braket}
\usepackage{enumerate}
\usepackage[normalem]{ulem}
\usepackage[usenames,dvipsnames]{xcolor}
\usepackage{multirow}
\usepackage{mathtools}

\definecolor{orange}{rgb}{1,0.5,0}

\begin{document}

\title{Error Resilience and Unitary 2-Designs}

\author{Zheng-Hang Sun}
\affiliation{Theoretical Physics III, Center for Electronic Correlations and Magnetism,
Institute of Physics, University of Augsburg, 86135 Augsburg, Germany}

\author{Fabian Ballar Trigueros~\orcidlink{0000-0001-6395-8767}}
\email{fabian.ballar@uni-a.de}
\affiliation{Theoretical Physics III, Center for Electronic Correlations and Magnetism,
Institute of Physics, University of Augsburg, 86135 Augsburg, Germany}

\author{Xhek Turkeshi~\orcidlink{0000-0003-1093-3771}}
 \email{turkeshi@thp.uni-koeln.de}
\affiliation{Institut f\"ur Theoretische Physik, Universit\"at zu K\"oln, Z\"ulpicher Strasse 77, 50937 K\"oln, Germany}
\affiliation{Institute of Quantum Control, Peter Grünberg Institut (PGI-8),
Forschungszentrum Jülich GmbH, 52425 Jülich, Germany}

\author{Piotr Sierant~\orcidlink{0000-0001-9219-7274}}
\email{piotr.sierant@bsc.es}
\affiliation{Barcelona Supercomputing Center, Plaça Eusebi G\"uell, 1-3 08034, Barcelona, Spain}

\begin{abstract}
\noindent 
Encoding-error-decoding circuits provide a setting for studying protection of quantum information against errors. In these circuits, an encoding unitary is followed by an error layer, decoding by the inverse unitary, and ancilla postselection.
For encoders drawn from unitary 2-design ensembles, these circuits exhibit an error-protecting phase  below a threshold error strength in which the fidelity between the input and recovered logical states approaches unity.
Here, we show how departures from Haar second-moment statistics of the encoder can destabilize the error-protecting phase and derive a universal inequality linking recovery fidelity to the second unitary frame potential, which quantifies the ensemble’s proximity to a unitary 2-design.
This connection is exemplified by matchgate circuits doped with a tunable number of non-Gaussian gates. 
We analytically show that undoped circuits lack a stable error-protecting phase in the thermodynamic limit and that increasing the doping restores the error protection mechanism as the encoder ensemble approaches an approximate unitary 2-design.
Further examples of Hamiltonian encoders, including an experimentally motivated all-to-all spin model, follow the same mechanism.
Our results identify recovery fidelity as a diagnostic of departures from unitary 2-design behavior in quantum many-body systems.
\end{abstract}
\maketitle

\emph{Introduction.---}
Understanding how errors degrade encoded quantum information is central to realizing practical quantum technologies~\cite{Preskill2018quantumcomputingin}. In the quantum error-correction framework, one naturally considers an encoding map that embeds $k$ logical qubits into $N$ physical qubits, a noise channel, and a decoding operation to reconstruct the logical state~\cite{PhysRevA.55.900,PhysRevLett.94.180501,RevModPhys.87.307,Shor1995memory,Steane1996codes}. In general, tackling this problem is hard because it depends on the choice of a good decoder. An alternative strategy to explore the effect of noise on quantum information is offered by encoding-error-decoding (EED) protocols. In this framework, $k$ system qubits are encoded together with $N-k$ ancillas through an $N$-qubit unitary $U$ and then subjected to a layer of errors. The decoder is chosen simply as $U^\dagger$, followed by syndrome measurements with postselection allowed. In the scaling limit $N\to\infty$ at fixed code rate $r=k/N$, an error-resilience phase transition occurs as the error strength is varied~\cite{PhysRevLett.132.140401,sierant2026}. Below a critical threshold, the input state on $k$ qubits is recovered (error-protecting phase); above it, the encoded information washes out (error-vulnerable phase). While EED protocols differ from quantum error correction, the persistence of faithful recovery at finite error strength is reminiscent of the threshold theorem for fault-tolerant quantum computation~\cite{1999quant.ph..6129A}.

Beyond this analogy, the EED framework has become a useful lens for circuit-inversion benchmarks on quantum hardware~\cite{PhysRevLett.129.150502}, emergent unitary designs~\cite{bnld-2chd}, and resource-theoretic properties such as nonstabilizerness~\cite{Niroula:2024wp,cbjz-x45n,sierant2026,Feng2026resource}. The transition is also experimentally relevant, with signatures recently observed on a superconducting quantum processor~\cite{Dallas26exp}. Error resilience has been established for Haar-random and random Clifford encoders~\cite{PhysRevLett.132.140401,sierant2026}; for qubits, the latter form a unitary $3$-design, matching the Haar average up to third moments~\cite{Zhu2017clifford}. Nevertheless, the general conditions an encoder ensemble must satisfy to sustain a stable error-protecting phase remain unclear.

In this work, we clarify the relation between error resilience and the randomness of the encoder ensemble. To that end, we first derive a universal inequality connecting the recovery fidelity in the EED setup to the second unitary frame potential. We then study the EED with encoders taken as doped matchgate circuits~\cite{10.1098/rspa.2008.0189,2001quant.ph..8033K,Wan:2023ty,PhysRevA.65.032325,2025arXiv250524212B}, that approach the 2-design behavior when the number of non-Gaussian gates is sufficiently increased~\cite{dopeddesigns}. We analytically show that for matchgates without doping, the error-protecting regime shrinks with increasing system size and disappears in the thermodynamic limit. The regime becomes more stable by doping matchgate circuits with non-Gaussian gates. We further substantiate this connection using Gaussian unitary ensemble (GUE) Hamiltonian dynamics~\cite{Cotler:2017va}, where approximate 2-designs can be achieved and an error-resilience phase transition is observed. In contrast, the dynamics of the clean non-integrable Ising model
~\cite{PhysRevLett.106.050405,PhysRevLett.111.127205,PhysRevX.14.031014}, remains insufficiently random at the level of the second unitary moment~\cite{Roberts:2017tk}, and correspondingly does not support a stable error-protecting phase. We also study an experimentally feasible disordered all-to-all spin model, in which the recovery of error resilience closely tracks the emergence of 2-designs. 
\begin{figure}[]
	\centering
	\includegraphics[width=1\linewidth]{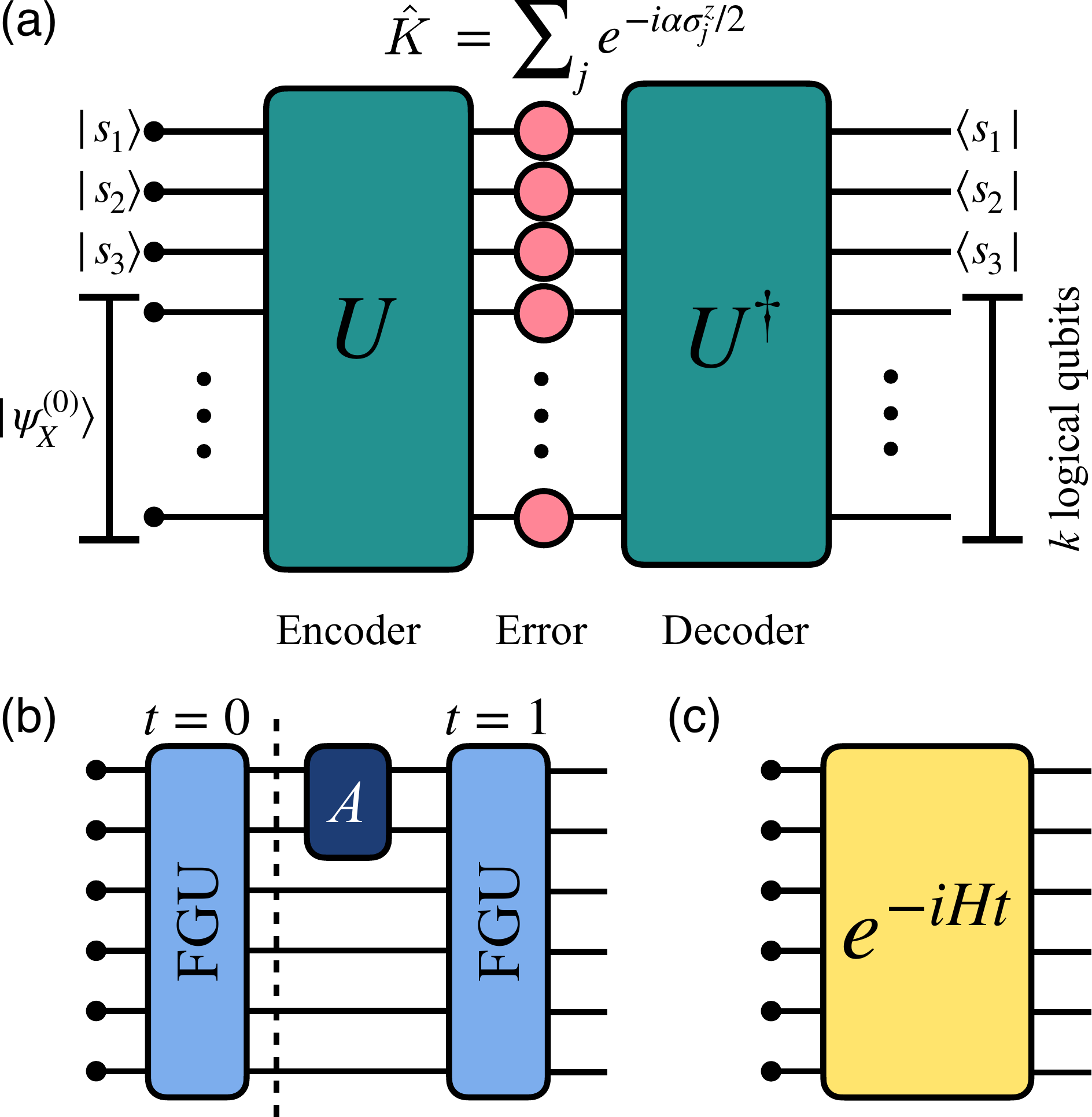}\\
	\caption{(a) Schematic illustration of the encoding-error-decoding quantum evolutions comprised of the encoder $U$, the local error layer $\hat{K}$, and the decoder $U^{\dagger}$. The $N$-qubit system consists of $k$ logical qubits with the initial state $|\psi_{X}^{(0)}\rangle$ and $N-k$ ancilla qubits which are projected onto their initial states after the decoder $U^{\dagger}$. (b) Doped fermionic Gaussian unitary (FGU) circuit considered as one choice of the encoder $U$, where non-Gaussian gates $A$ are inserted into otherwise Gaussian matchgate circuits. The pure matchgate limit corresponds to $t=0$. (c) Hamiltonian time evolution $U=e^{-iHt}$, considered as another choice of the encoder.}\label{fig1}
\end{figure}
Taken together, our results demonstrate that the error-resilience transition in the EED protocols 
reflects the second-moment randomness.

\begin{figure*}[]
	\centering
	\includegraphics[width=1\linewidth]{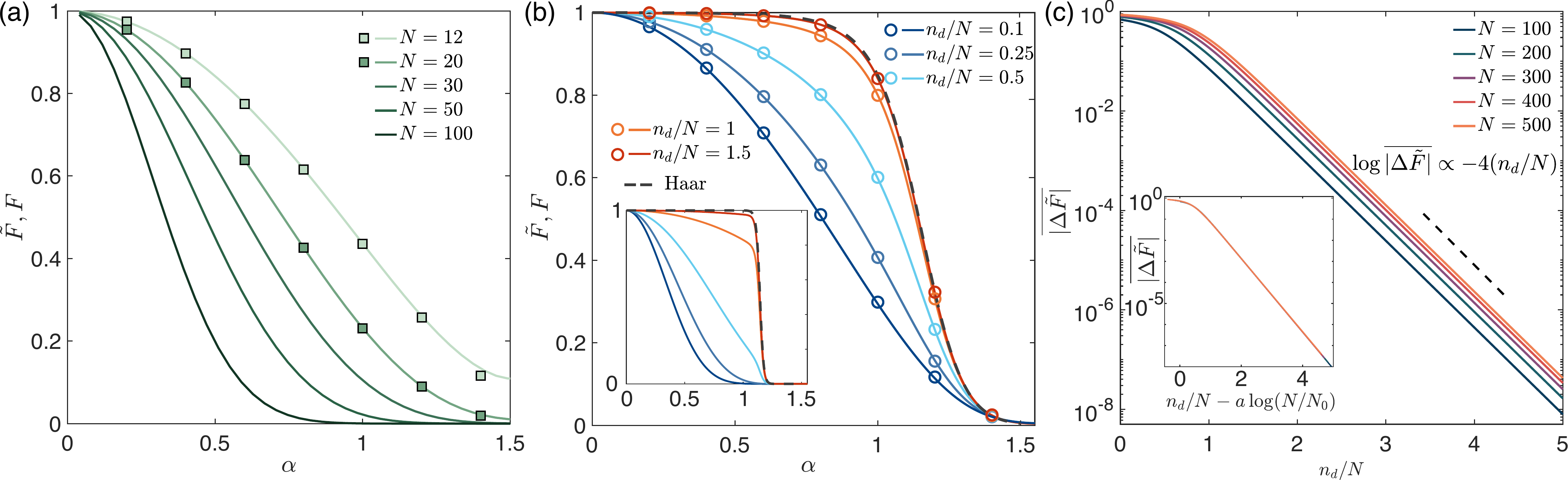}\\
	\caption{(a) For EED matchgate circuits without non-Gaussian doping gates, the fidelity versus error strength $\alpha$. Markers show the fidelity $F$ in Eq.~(\ref{fidelity}), while solid lines show the annealed fidelity $\tilde{F}$ in Eq.~(\ref{eq3}). (b) For doped matchgate circuits with $N=20$, the fidelity versus $\alpha$; the inset shows $N=100$. Markers and lines show $F$ and $\tilde F$ from Eq.~(\ref{eq6}), and dashed lines show the Haar result. (c) The value of $\overline{|\Delta\tilde{F}|}$ as a function of the number of non-Gaussian doping gates $n_{d}$ with different system sizes $N$. The inset of (c) shows a data collapse of $\overline{|\Delta\tilde{F}|}$ with the form of the function $\overline{|\Delta\tilde{F}|} = f[n_{d}/N - a\log(N/N_{0})]$ with $a\simeq 0.26$ and $N_{0}\simeq 100$.  }\label{fig2}
\end{figure*}

\emph{Encoding-error-decoding quantum evolutions and state designs.---}
We start by presenting the general EED protocol, which can be framed as a quantum evolution of $N$ qubits [Fig.~\ref{fig1}(a)]. The system is divided into a subsystem $X$ of $k$ logical qubits and its complement $\bar{X}$ of $N-k$ ancillas, initialized in a computational-basis product state $|\psi_{\mathrm{in}}\rangle=|\psi^{\mathrm{in}}_{X}\rangle\otimes|\psi^{\mathrm{in}}_{\bar{X}}\rangle=\bigotimes_{i=1}^{N}|s_i\rangle$, with $s_i\in\{0,1\}$ and $\rho_{\mathrm{in}}=|\psi_{\mathrm{in}}\rangle\langle\psi_{\mathrm{in}}|$. The state is encoded by a unitary $U$ drawn from an ensemble $\mathcal{E}$, giving $\rho_U=U\rho_{\mathrm{in}}U^{\dagger}$, and then subjected to a layer of local errors $\mathcal{K}_\alpha=\mathop{\bigotimes}_{j=1}^{N}\mathcal{K}_{j,\alpha}$ of tunable strength $\alpha$, which corrupt the encoded information. To retrieve it, a decoder $\mathcal{D}$ is applied to $\mathcal{K}_\alpha(\rho_U)$, together with syndrome measurements on the ancillas. EED protocols fix the decoder to the inverse encoding unitary, $\mathcal{D}(\rho)=U^{\dagger}\rho U$, and postselect the ancilla measurements on their initial state. The output state $\rho_{\mathrm{out}}=U^{\dagger}\mathcal{K}_\alpha(\rho_U)U$ thus yields the logical state $\rho_X=\langle\psi^{\mathrm{in}}_{\bar{X}}|\rho_{\mathrm{out}}|\psi^{\mathrm{in}}_{\bar{X}}\rangle/p$, where $p=\mathrm{tr}(\langle\psi^{\mathrm{in}}_{\bar{X}}|\rho_{\mathrm{out}}|\psi^{\mathrm{in}}_{\bar{X}}\rangle)$ is the postselection probability. To quantify error resilience, we use the EED fidelity
\begin{equation}
F=\langle\psi^{\mathrm{in}}_{X}|\rho_{X}|\psi^{\mathrm{in}}_{X}\rangle=\frac{\langle\psi_{\mathrm{in}}|\rho_{\mathrm{out}}|\psi_{\mathrm{in}}\rangle}{p},
\label{fidelity}
\end{equation}
i.e., the probability that the logical state is recovered, conditioned on the ancillas returning to their initial state. Related scrambling diagnostics are based on quantum channels and recovery protocols~\cite{Hosur2016chaos,Yoshida2019scrambling}.

Throughout this work, we consider coherent errors, $\mathcal{K}_\alpha(\rho)=\hat{K}\rho\hat{K}^{\dagger}$ with $\hat{K}=\exp[-i(\alpha/2)\sum_{j=1}^{N}\sigma_{j}^{z}]$, which arise in current quantum devices from calibration imperfections. The output state is then pure, $\rho_{\mathrm{out}}=|\psi_{\mathrm{out}}\rangle\langle\psi_{\mathrm{out}}|$ with $|\psi_{\mathrm{out}}\rangle=U^{\dagger}\hat{K}U|\psi_{\mathrm{in}}\rangle$. For Haar-random encoders at code rate $r=k/N=1/2$, $F$ undergoes an error-resilience phase transition at $\alpha_c^{\mathrm{Haar}}=2\arccos(2^{-1/4})\approx1.14$: for $\alpha<\alpha_c^{\mathrm{Haar}}$ ($\alpha>\alpha_c^{\mathrm{Haar}}$), $F\to1$ ($F\to0$) as $N\to\infty$, defining the error-protecting (error-vulnerable) phase~\cite{PhysRevLett.132.140401}. The transition stems from a competition between two contributions to $p$: the probability that the encoded state is left invariant by $\hat{K}$, which decays as $\cos^{2N}(\alpha/2)$, and the probability that it is rotated into one of the $2^k-1$ orthogonal logical states, which scales as $2^{-N(1-r)}$. Balancing the two gives $\cos^{2}(\alpha_c^{\mathrm{Haar}}/2)=2^{-(1-r)}$, which reduces to the value above at $r=1/2$.

Since the encoder is drawn from an ensemble $\mathcal{E}$, error resilience is characterized by the average fidelity $\overline{F}=\mathbb{E}_{U\sim\mathcal{E}}[F]$. 
Computing this quenched average analytically is generally hard since $F$ is a ratio of two $U$-dependent quantities, and its ensemble average does not reduce to a finite number of moments of $U$. To circumvent this problem, one can study the annealed fidelity $\tilde{F}=\mathbb{E}_{U\sim\mathcal{E}}[\langle\psi_{\mathrm{in}}|\rho_{\mathrm{out}}|\psi_{\mathrm{in}}\rangle]/\mathbb{E}_{U\sim\mathcal{E}}[p]$, in which numerator and denominator are averaged separately, and depend only on the second unitary moment.
In several cases, both are self-averaging, and $\tilde{F}$, which is tractable analytically, provides quantitative predictions for the quenched average as well~\cite{PhysRevLett.132.140401}. Using replica boundary states~\cite{PhysRevLett.132.140401}, the annealed fidelity reads
\begin{equation}
\tilde{F}=\frac{\mathrm{Tr}[\Phi_{\mathcal{E}}^{(2)}(\hat{K}\otimes\hat{K}^{\dagger})B_{\mathrm{num}}^{F}]}{\mathrm{Tr}[\Phi_{\mathcal{E}}^{(2)}(\hat{K}\otimes\hat{K}^{\dagger})B_{\mathrm{den}}^{F}]},
\label{eq3}
\end{equation}
where $B_{\mathrm{num}}^{F}=\rho_{\mathrm{in}}^{\otimes2}$, $B_{\mathrm{den}}^{F}=\rho_{\mathrm{in}}\otimes(\mathbb{I}_{X}\otimes|\psi^{\mathrm{in}}_{\bar{X}}\rangle\langle\psi^{\mathrm{in}}_{\bar{X}}|)\,T_{(1,2)}$, $T_{(1,2)}$ is the permutation swapping the two replicas, and $\Phi_{\mathcal{E}}^{(2)}(\hat{\mathcal{O}})\equiv\mathbb{E}_{U\sim\mathcal{E}}[(U^{\dagger})^{\otimes2}\hat{\mathcal{O}}\,U^{\otimes2}]$ is the second-moment twirl~\cite{Dankert2009designs}. Since $\tilde{F}$ depends on $\mathcal{E}$ only through $\Phi_{\mathcal{E}}^{(2)}$, it coincides with its Haar value for any unitary $2$-design.

Our goal is to relate error resilience to the randomness of the encoder ensemble. A natural way to quantify the latter is to compare the moments of $\mathcal{E}$ with those of the maximally random Haar ensemble. At order $t$, this comparison is captured by the frame potential $\mathcal{F}^{(t)}_{\mathcal{E}}=\mathbb{E}_{U,V\sim\mathcal{E}}|\mathrm{Tr}(U^{\dagger}V)|^{2t}$~\cite{Gross2007designs,Roberts:2017tk}, which satisfies $\mathcal{F}^{(t)}_{\mathcal{E}}\geq\mathcal{F}^{(t)}_{\mathrm{Haar}}$, with equality if and only if $\mathcal{E}$ is a unitary $t$-design. Since the annealed fidelity~\eqref{eq3} involves only second moments, we focus on $t=2$ and define the design excess $\Delta_{\mathcal{E}}\equiv\mathcal{F}^{(2)}_{\mathcal{E}}-\mathcal{F}^{(2)}_{\mathrm{Haar}}$, with $\mathcal{F}^{(2)}_{\mathrm{Haar}}=2$ for the Haar measure on $U(d)$, $d=2^N$. Equivalently, $\Delta_{\mathcal{E}}$ is the squared Hilbert--Schmidt distance between the second-moment operators of $\mathcal{E}$ and of the Haar ensemble. We now show that $\Delta_{\mathcal{E}}$ controls how far the annealed fidelity can deviate from its Haar value.

In the EED setup of Fig.~\ref{fig1}(a), with $1\le k\leq N$ and a fixed unitary error layer $\hat{K}$, any encoder ensemble $\mathcal{E}$ with average postselection probability $\bar{p}_{\mathcal{E}}\equiv\mathbb{E}_{U\sim\mathcal{E}}[p]>0$ satisfies
\begin{equation}
\Delta_{\mathcal{E}}\geq\frac{\bar{p}_{\mathcal{E}}^{\,2}\,|\tilde{F}-\tilde{F}_{\mathrm{Haar}}|^{2}}{d^{2}\left(1+D\tilde{F}_{\mathrm{Haar}}^{\,2}-2\tilde{F}_{\mathrm{Haar}}\right)},
\label{eq:eed_witness}
\end{equation}
where $D=2^k$ and $\tilde{F}_{\mathrm{Haar}}=(c+d)(d-1)/[(c+d)(d-1)+(D-1)(d^{2}-c)]$, with $c=|\mathrm{Tr}\hat{K}|^{2}$, is the annealed fidelity of Haar-random encoders at the same $N$, $k$, $\hat{K}$, with the same input state and postselection~\cite{PhysRevLett.132.140401}. Exact $2$-designs ($\Delta_{\mathcal{E}}=0$) thus reproduce $\tilde{F}_{\mathrm{Haar}}$, and any deviation from it requires a finite design excess. The same bound holds for encoders restricted to a symmetry class, i.e., to a compact subgroup $G\subset U(d)$, with $\mathcal{F}^{(2)}_{\mathrm{Haar}}$ and $\tilde{F}_{\mathrm{Haar}}$ replaced by their values for the Haar measure on $G$.

To illustrate how this connection manifests in concrete encoders, we study two classes of unitaries $U$: doped matchgate circuits and Hamiltonian time evolution, with input states $|\psi_{\mathrm{in}}\rangle=|0\rangle^{\otimes N}$ and the N\'{e}el state, respectively.
\emph{
Doped matchgate circuits.---} 
We first consider an undoped random matchgate encoder $U_{\mathrm{M}}$ [see Fig.~\ref{fig1}(b) with $t=0$]. Via the Jordan--Wigner transformation, parity-preserving matchgates correspond to fermionic Gaussian unitaries $U_{\mathrm{M}}=\exp[\frac{1}{4}\sum_{\mu,\nu=1}^{2N}G_{\mu\nu}\gamma_\mu\gamma_\nu]$, where $G=-G^{T}$ is real and $\gamma_{2j-1}=(\prod_{l<j}Z_l)X_j$, $\gamma_{2j}=(\prod_{l<j}Z_l)Y_j$ are Majorana operators, satisfying $\{\gamma_\mu,\gamma_\nu\}=2\delta_{\mu\nu}$~\cite{PhysRevA.65.032325,Casas2026matchgate,Bravyi2005FLO}. Their Majorana action lies in $SO(2N)$, which can be sampled using Haar-random active fermionic-linear-optics (FLO) circuits~\cite{2025arXiv250524212B}. The second-moment average used below is Haar over the full orthogonal Gaussian ensemble $O(2N)$, obtained by composing these draws with a fixed improper Majorana reflection with probability $1/2$~\cite{Wan:2023ty}.

The annealed fidelity~(\ref{eq3}) is exactly computable from the two-copy twirl ${\Phi_{\mathrm{M}}^{(2)}(\hat{\mathcal{O}})\equiv\mathbb{E}_{U_{\mathrm{M}}}[(U_{\mathrm{M}}^{\dagger})^{\otimes2}\hat{\mathcal{O}}\,U_{\mathrm{M}}^{\otimes2}]}$~\cite{Sierant2026commutant,Braccia2026commutant,Lastres2026commutants}. This is the orthogonal projector onto the two-copy commutant of the $O(2N)$ Gaussian unitaries, which is spanned by $2N+1$ operators~\cite{Wan:2023ty}, so that
\begin{equation}
\Phi_{\mathrm{M}}^{(2)}(\hat{\mathcal{O}})=\sum_{\ell=0}^{2N}\mathrm{Tr}\big(\Upsilon_{\ell}\,\hat{\mathcal{O}}\big)\,\Upsilon_{\ell},\quad
\Upsilon_{\ell}=\frac{\sum_{|S|=\ell}\gamma_{S}\otimes\gamma_{S}}{2^N\sqrt{\binom{2N}{\ell}}},
\label{eq1}
\end{equation}
where $\gamma_{S}=\prod_{\mu\in S}\gamma_{\mu}$ for $S\subseteq\{1,\dots,2N\}$, and $\mathrm{Tr}(\Upsilon_{\ell}\Upsilon_{\ell'})=\delta_{\ell\ell'}$. Since $\hat{K}=\prod_{j}[\cos(\alpha/2)-\sin(\alpha/2)\gamma_{2j-1}\gamma_{2j}]$, only even shells contribute, $\Phi_{\mathrm{M}}^{(2)}(\hat{K}\otimes\hat{K}^{\dagger})=d\sum_{m=0}^{N}s_{2m}\Upsilon_{2m}$, and Eq.~(\ref{eq3}) becomes
\begin{equation}
\tilde{F}_{\mathrm{M}}=\frac{\sum_{m=0}^{N}s_{2m}N_{2m}}{\sum_{m=0}^{N}s_{2m}D_{2m}},
\label{eq:Fmatch}
\end{equation}
with $N_{2m}\equiv\mathrm{Tr}(\Upsilon_{2m}B^{F}_{\mathrm{num}})$ and $D_{2m}\equiv\mathrm{Tr}(\Upsilon_{2m}B^{F}_{\mathrm{den}})$. As detailed in the Supplementary Material, for logical and ancilla qubits initialized in the computational basis we find
\begin{align}
D_{2m}&=2^{-N}\binom{2N}{2m}^{-1/2}(-1)^{m}\sum_{m'=0}^{\min(m,N-k)}\binom{N-k}{m'}\binom{2k}{2(m-m')},\nonumber\\
s_{2m}&=\binom{2N}{2m}^{-1/2}\binom{N}{m}(-1)^{m}\Big(\cos^{2}\frac{\alpha}{2}\Big)^{N-m}\Big(\sin^{2}\frac{\alpha}{2}\Big)^{m},\nonumber\\
N_{2m}&=2^{-N}\binom{2N}{2m}^{-1/2}\binom{N}{m}(-1)^{m},
\label{eq:coeffs}
\end{align}
where we notice that the signs $(-1)^m$ cancel in Eq.~(\ref{eq:Fmatch}), so that all terms are positive.

Figure~\ref{fig2}(a) shows $\tilde{F}_{\mathrm{M}}$ as a function of the error strength $\alpha$ for code rate $r=k/N=1/2$ and different system sizes $N$ (solid lines). Here and in the following, whenever the encoder is random, numerical data for $F$ (markers) are averaged over its realizations, i.e., they give the quenched average $\overline{F}=\mathbb{E}_{U\sim\mathcal{E}}[F]$, obtained here from FLO simulations. The annealed and quenched fidelities agree well, with $|\overline{F}-\tilde{F}_{\mathrm{M}}|<10^{-2}$ (see Supplementary Material). With increasing $N$, the fidelity drops at increasingly small $\alpha$, and Eq.~(\ref{eq:Fmatch}) gives $\tilde{F}_{\mathrm{M}}\to0$ as $N\to\infty$ for every fixed $0<\alpha<\pi$ (see Supplementary Material). Random matchgate encoders therefore do not support a stable error-protecting phase.

Next, we study the EED setup with doped matchgate circuits $U_{\mathrm{D}}$ in Fig.~\ref{fig1}(b), i.e., $U_{\mathrm{D}} = [ \prod_{i=1}^{n_{d}} (U_{\mathrm{M}}^{(i)} A) ] U_{\mathrm{M}}^{(0)}$, with $U_{\mathrm{M}}^{(i)}$ ($i=0,1,\dots,n_{d}$) being independent draws from the matchgate ensemble and $A = \exp[-i( \pi/4) Z_{1}Z_{2}]$ referring to the non-Gaussian doping gate~\cite{Brod2011matchgates}. Increasing the density of non-Gaussian gates breaks the fermionic Gaussian structure of the matchgate circuits~\cite{Sierant2026fermionicmagic,Leone2026fermionicentropy, Tarabunga26covariance} and drives the circuit ensemble toward an approximate symmetry-resolved unitary 2-design~\cite{dopeddesigns}.
By using the doped matchgate twirl in Ref.~\cite{dopeddesigns} (also see Supplementary Material), the annealed fidelity for the EED circuits can be analytically expressed as
\begin{equation}
\tilde{F}_{\mathrm{D}} = \frac{\sum_{r,m = 0}^{N} N_{2r} (\Gamma^{n_{d}})_{r,m}s_{2m} }{\sum_{r,m = 0}^{N} D_{2r} (\Gamma^{n_{d}})_{r,m}s_{2m}},
\label{eq6}
\end{equation}
which reduces to Eq.~(\ref{eq:Fmatch}) for $n_d=0$. Here, $s_{2m}$, $N_{2m}$, and $D_{2m}$ are defined in Eq.~(\ref{eq:coeffs}), $n_{d}$ is the number of non-Gaussian doping gates, and $\Gamma_{r,m}=C_{2r,2m}$ is the even-shell transfer matrix, cf. Supplementary Material for details.

Based on the analytical expression of $\tilde{F}$ in Eq.~(\ref{eq6}), we plot $\tilde{F}$ as a function of error strength $\alpha$ with $N=20$ and different $n_{d}/N$ in Fig.~\ref{fig2}(b). With increasing $n_{d}$, in the regime $\alpha<\alpha_{c}^{\mathrm{Haar}}$, $\tilde{F}$ increases toward its large-doping limit, indicating enhanced error resilience. The same tendency is observed for $N=100$ [inset of Fig.~\ref{fig2}(b)], consistent with the stabilization of an error-protecting phase in the thermodynamic limit when $n_{d}$ grows sufficiently rapidly with $N$.

We quantify this restoration of error resilience by the protected-window average of the fidelity deviation from the Haar result, defined as $\overline{|\Delta\tilde F|} = \frac{1}{\alpha_{c}^{\mathrm{Haar}}} \int_{0}^{\alpha_{c}^{\mathrm{Haar}}} \mathrm{d} \alpha\, |\tilde{F}_{\mathrm{D}}(\alpha) - \tilde{F}_{\mathrm{Haar}}(\alpha) |$. Figure~\ref{fig2}(c) shows that $\overline{|\Delta\tilde F|}$ decays approximately as $e^{-4\tau}$ in the late-doping regime shown, with $\tau=n_d/N$. Remarkably, this is the same late-doping slope as the excess second unitary frame potential above the symmetry-constrained Haar value~\cite{dopeddesigns}. Moreover, in the inset of Fig.~\ref{fig2}(c), we show a data collapse following the form $\overline{|\Delta\tilde F|} \simeq f[n_{d}/N - a \log (N/N_{0})]$, indicating that a fixed averaged fidelity deviation $\overline{|\Delta\tilde F|}$ can be reached on the scale $n_d\sim N\log N$. This is particularly suggestive because Ref.~\cite{dopeddesigns} establishes $\Omega(N\log N)$ as a lower bound for approximate unitary $2$-design formation within this symmetry class in the same doped-matchgate architecture. An analytic explanation of the common slope is given in the End Matter.

\emph{Hamiltonian dynamics.---}
Finally, we study the EED protocol with encoders generated by time-independent Hamiltonian dynamics, $U = \exp(-iHt)$ [Fig.~\ref{fig1}(c)]. 

\begin{figure}[]
	\centering
	\includegraphics[width=1\linewidth]{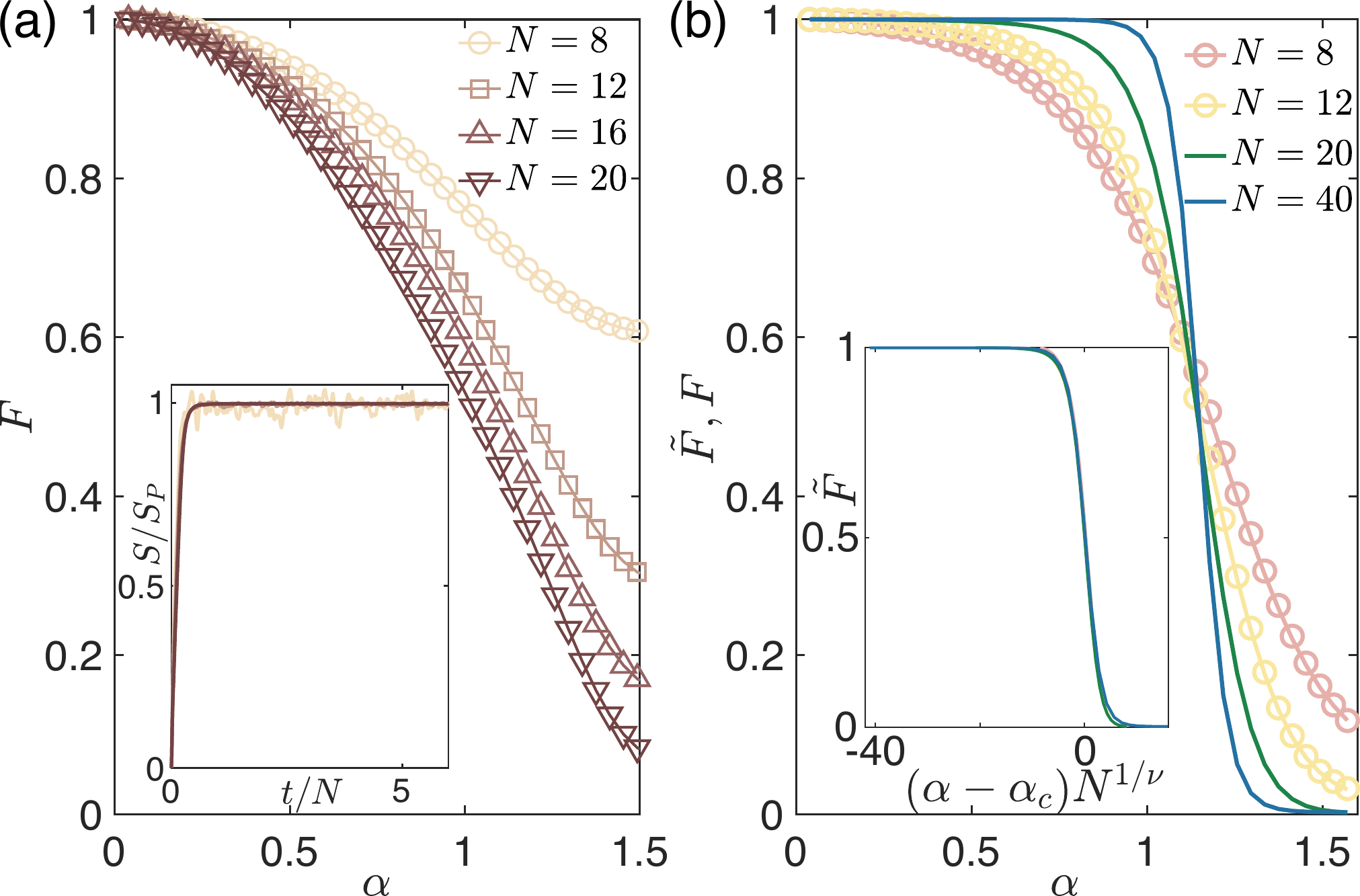}\\
	\caption{(a) Fidelity $F$, Eq.~(\ref{fidelity}), versus error strength $\alpha$ for the EED quantum evolution with the Ising Hamiltonian dynamics as the encoder, for different system sizes $N$. The inset shows the dynamics of the half-chain entanglement entropy. (b) Same as (a), but with GUE Hamiltonian dynamics as the encoder. Markers show the quenched average $\overline{F}$ of the fidelity~(\ref{fidelity}), while solid lines show the annealed fidelity $\tilde{F}_{\mathrm{GUE}}$, Eq.~(\ref{eqf_gue}). The inset shows the data collapse of $\tilde{F}_{\mathrm{GUE}}$ as a function of $(\alpha-\alpha_{c})N^{1/\nu}$.}\label{fig3}
\end{figure}

We first consider a non-integrable quantum Ising model~\cite{PhysRevLett.111.127205,Haldar21sign} $H_{\text{Ising}} = -\sum_{i=1}^{N-1} X_{i}X_{i+1} - \sum_{i=1}^{N-2}X_{i}X_{i+2} - \sum_{i=1}^{N} Z_{i}$. The N\'{e}el state has zero energy, $\langle H_{\text{Ising}}\rangle=\mathrm{Tr}(H_{\text{Ising}})/d=0$, and thus thermalizes to infinite temperature: its half-chain entanglement entropy $S=-{\rm Tr}\rho_A\log\rho_A$, with $\rho_A$ the reduced density matrix of half of the chain, rapidly approaches the Page value $S_{P}\simeq (N\log 2 - 1)/2$~\cite{PhysRevLett.71.1291,Bianchi2022entanglement}, a clear signature of quantum thermalization~\cite{PhysRevLett.128.160502, Polkovnikov11coll,DAlessio2016thermalization,Kaufman2016thermalization,Neill2016thermalization} [see the inset of Fig.~\ref{fig3}(a)].

For the EED quantum evolution with the encoding unitary $U= \exp(-iH_{\text{Ising}}t)$ and $t=6N$, we numerically calculate the fidelity~(\ref{fidelity}), see Fig.~\ref{fig3}(a). Unlike the previous examples, the encoder is now a single deterministic unitary, and no ensemble average is involved. 
Similar to the case of undoped matchgate circuits, there is no stable error-protecting phase, in which $F\to1$ with increasing $N$, and no signature of the error-resilience phase transition observed for Haar-random encoders. 
Thermalization of the encoded state, even at infinite temperature, is therefore not sufficient for error resilience. While the numerator of $F$ depends only on the evolved state $U|\psi_{\rm in}\rangle$, the postselection probability $p$ depends on how the decoder $U^\dagger$ acts on the error-affected state, and thus probes the randomness of $U$ itself. At this level, local Hamiltonian dynamics is insufficiently random: even ensembles of constant-local Hamiltonians fail to form 2-designs at any evolution time~\cite{cui2025randomunitarieshamiltoniandynamics}. In contrast, for the ensemble generated by the unitary dynamics of a GUE Hamiltonian, approximate 2-designs can emerge~\cite{Cotler:2017va}.

To further clarify the role of approximate 2-designs in the emergence of error resilience, we consider the encoding unitary $U = \exp(-i H_{\text{GUE}} t )$ at $t=\sqrt{d}$, an appropriate time for approximate 2-designs~\cite{Cotler:2017va}. For this setup, the annealed fidelity can be derived analytically as
\begin{equation}
\tilde{F}_{\mathrm{GUE}} = \frac{\sum_{\sigma,\tau\in S_{4}}
\mathrm{Wg}_{d}(\sigma,\tau)
\left[
\prod_{c\in \mathrm{cyc}(\gamma\sigma)}
g_{\mathrm{N}}(c)
\right]
\mathcal{R}_{\tau}}{\sum_{\sigma,\tau\in S_{4}}
\mathrm{Wg}_{d}(\sigma,\tau)
\left[
\prod_{c\in \mathrm{cyc}(\gamma\sigma)}
g_{\mathrm{D}}(c)
\right]
\mathcal{R}_{\tau}},
\label{eqf_gue}
\end{equation}
where $\mathrm{Wg}_{d}(\sigma,\tau)$ is the $S_{4}$ Weingarten matrix~\cite{Collins2006integration}, $\mathcal{R}_{\tau}$, as a function of $t$, is the GUE spectral factor from Ref.~\cite{Cotler:2017va}, and $\gamma = (1234)$. The numerator and denominator differ only in the cycle factors $g_{\mathrm{N}}(c)$ and $g_{\mathrm{D}}(c)$. The details of the derivation and the expressions of $g_{\mathrm{N}}(c)$, $g_{\mathrm{D}}(c)$, and $\mathcal{R}_{\tau}$ are provided in the Supplementary Material.

In Fig.~\ref{fig3}(b), we plot the fidelity for the EED with GUE Hamiltonian dynamics. As in the Haar-random case~\cite{PhysRevLett.132.140401}, we observe a clear signature of a phase transition between the error-protecting and error-vulnerable phases, where the fidelity tends to 1 and 0, respectively. A data collapse of the form $\tilde{F}_{\mathrm{GUE}} = f[(\alpha-\alpha_{c})N^{1/\nu}]$ gives the transition point $\alpha_{c} \simeq 1.14$, in agreement with $\alpha_c^{\mathrm{Haar}}$, and the critical exponent $\nu\simeq 1.02$ [see the inset of Fig.~\ref{fig3}(b)].

\begin{figure}[]
	\centering
	\includegraphics[width=1\linewidth]{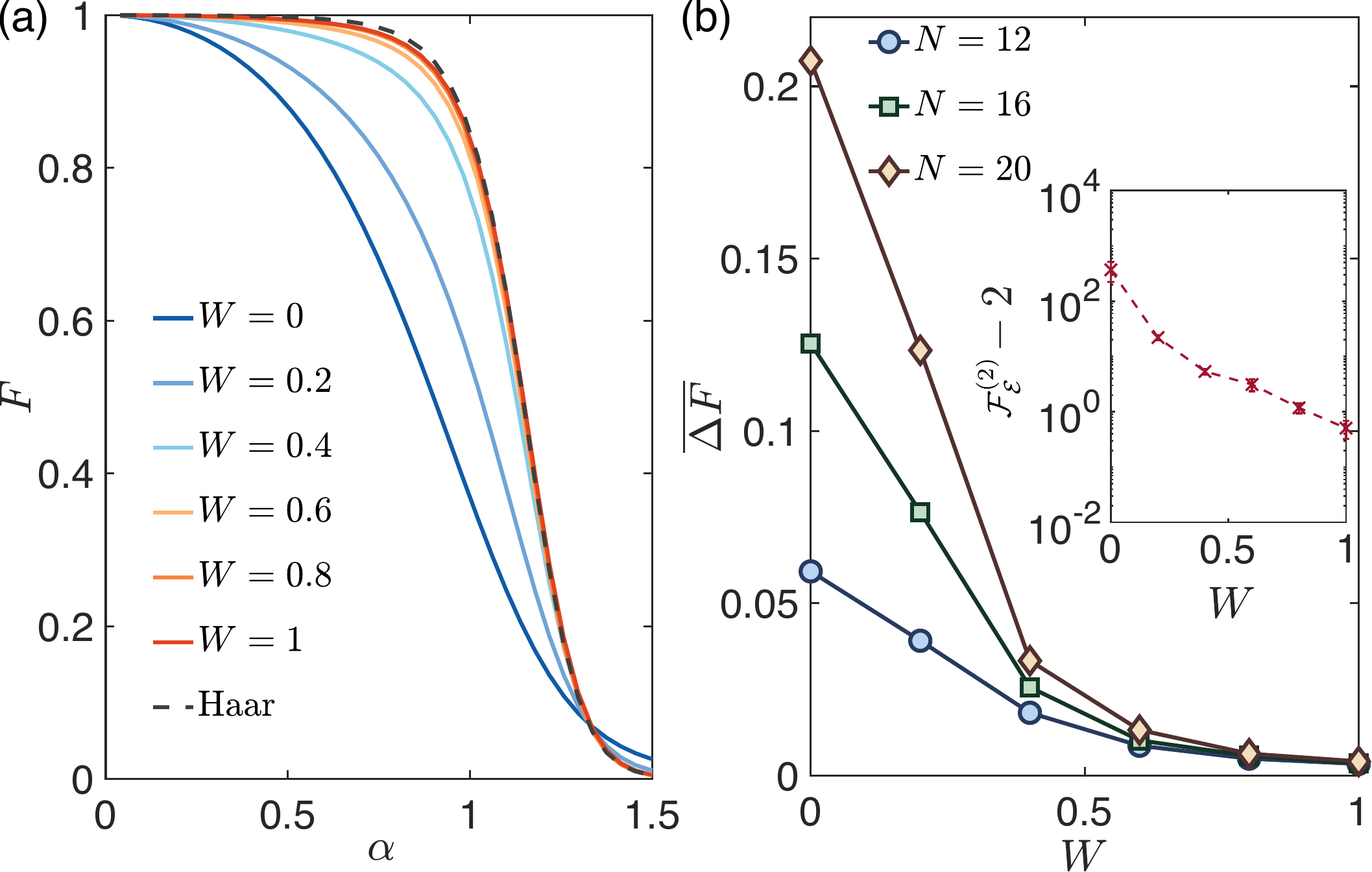}\\
	\caption{(a) Fidelity $F$, Eq.~(\ref{fidelity}), versus error strength $\alpha$ for the EED quantum evolution generated by Hamiltonian~(\ref{a2a_exp}), for different disorder strengths at $N=20$. Dashed lines show the Haar-random result from Ref.~\cite{PhysRevLett.132.140401}. (b) Averaged absolute fidelity deviation $\overline{|\Delta F|}$ versus $W$ for different system sizes. The inset shows the design excess $\Delta_{\mathcal{E}}=\mathcal{F}^{(2)}_{\mathcal{E}}-2$ versus $W$ at $N=12$. All results are averaged over 200 disorder realizations.}\label{fig4}
\end{figure}

Figure~\ref{fig3} separates conventional thermalization from Haar-like error resilience by contrasting a clean non-integrable Ising model with GUE Hamiltonian dynamics. We now tackle this contrast in an experimentally motivated all-to-all model. Specifically, we consider the Hamiltonian
\begin{equation}
H = \sum_{i<j}^{N} J_{ij}(W)(X_{i}X_{j} + Y_{i}Y_{j} ) + H_{F},
\label{a2a_exp}
\end{equation}
where $H_{F} = \sum_{i=1}^{N} (h_{i}^{x}X_{i} + h_{i}^{y}Y_{i} + h_{i}^{z}Z_{i})$ is the local-field Hamiltonian with $h_{i}^{x,y,z}$ independently drawn from the uniform distribution on $[-\sqrt{3}/2,\sqrt{3}/2]$, and the all-to-all coupling matrix is $J_{ij}(W) = -1/N + \eta_{ij}/\sqrt{N}$ with $\eta_{ij}$ independently drawn from the uniform distribution on $[-W,W]$. At $W=0$, the coupling matrix is uniform and the interaction term is permutation-symmetric. A finite $W$ breaks this symmetry, and increasing it makes the encoder more random. 
Related 2-local all-to-all $XY$ Hamiltonians have been implemented in atomic cavity QED~\cite{Sauerwein:2023tf,Davis2019spinexchange,Muniz2020cavity} and with superconducting qubits coupled to a common bus resonator~\cite{doi:10.1126/sciadv.aba4935}.

Figure~\ref{fig4}(a) shows the EED fidelity $F$ for the unitary dynamics of Eq.~(\ref{a2a_exp}) as the encoder, with $N=20$ and evolution time $t=10$. Increasing $W$ systematically shifts the fidelity curves towards the Haar result, as quantified by the rapid suppression of the averaged fidelity deviation $\overline{|\Delta F|} = \frac{1}{\alpha_{c}^{\mathrm{Haar}}} \int_{0}^{\alpha_{c}^{\mathrm{Haar}}} \mathrm{d} \alpha\, |F(\alpha) - F_{\mathrm{Haar}}(\alpha)|$ [see Fig.~\ref{fig4}(b)].
Crucially, this behavior is accompanied by the approach toward unitary 2-design behavior.
The inset of Fig.~\ref{fig4}(b) shows the design excess $\Delta_{\mathcal{E}}=\mathcal{F}^{(2)}_{\mathcal{E}}-2$ for the ensemble $\mathcal{E}$ generated by the unitary dynamics of Hamiltonian~(\ref{a2a_exp}) over disorder realizations. $\Delta_{\mathcal{E}}$ is suppressed by several orders of magnitude when $W$ increases, closely tracking the decay of $\overline{|\Delta F|}$.

\emph{Discussion.} -- In summary, we have connected the error resilience in the EED setup to second-moment randomness through a universal witness and complementary circuit and Hamiltonian examples. 
For purely Gaussian matchgate encoders, the error-protecting regime disappears in the thermodynamic limit at any finite coherent-error strength. Adding non-Gaussian gates restores it: the second-moment channel approaches its Haar form, and the EED fidelity converges to the Haar result exponentially in $n_d/N$.
A similar picture holds for Hamiltonian encoders. The dynamics of a clean non-integrable Ising model thermalizes without stabilizing an error-protecting regime, whereas GUE Hamiltonian dynamics reproduces the Haar-like fidelity transition. Finally, in an experimentally motivated random all-to-all $XY$ model, suppression of the excess second frame potential closely tracks recovery of Haar-like EED behavior.

Our inequality provides a general diagnostic of deviations from unitary $2$-design behavior, applicable to encoder ensembles generated by random quantum circuits or Hamiltonian dynamics~\cite{Emerson2003pseudorandom,Brandao2016designs,Onorati2017mixing,Zhou2026singlequench,Nandy2026unitarydesigns,SurNandy2026designs}.
More broadly, conventional diagnostics of thermalization and chaos, such as random-matrix level statistics~\cite{Chan2018spectral,Bertini2018spectral} and the growth of entanglement entropies~\cite{Nahum2017entanglement, Sierant23membrane}, cannot resolve the convergence of the second-moment channel to its Haar form~\cite{PhysRevX.15.011031,PhysRevX.14.041059}, and exact evaluation of unitary frame potentials is $\#\mathsf{P}$-hard~\cite{21vm-bz3t}. Estimating the EED fidelity, in contrast, requires only recovery and postselection measurements, avoiding reconstruction of the full second-moment channel. Implementing the EED protocol on programmable quantum simulators thus offers a practical route to witnessing deviations from unitary 2-design behavior, building on recent experimental advances in quantum pseudorandomness and scrambling~\cite{PhysRevLett.123.030502,Landsman:2019vn,cg3f-rggs, Choi2023randomstates,Garttner2017OTOC}.

\begin{acknowledgments}
\emph{Acknowledgments.} -- 
X.T. acknowledges support from DFG under Germany's Excellence Strategy – Cluster of Excellence Matter and Light for Quantum Computing (ML4Q) EXC 2004/2 – 390534769, and DFG Collaborative Research Center (CRC) 183 Project No. 277101999 - project B01, and DFG Emmy Noether Programme proposal ``Digital Quantum Matter Out-of-Equilibrium'' No. 560726973. 
P.S. acknowledges fellowship within the “Generación D” initiative, Red.es, Ministerio para la Transformación Digital y de la Función Pública, for talent attraction (C005/24-ED CV1), funded by the European Union NextGenerationEU funds, through PRTR.
\end{acknowledgments}

\emph{AI statement.---} We used Claude Opus 5.5 (Anthropic) to improve the writing and presentation of the manuscript. 
The authors reviewed all suggested changes and take full responsibility for the content and results of this work.

\section*{End matter.}
\subsection*{Second-moment witness and proof of Eq.~(\ref{eq:eed_witness}).}

\setcounter{equation}{0}
\renewcommand{\theequation}{E\arabic{equation}}

We use the same input, error layer, and ancilla projection
as in the main text. Set
$\rho_{\rm in}=\ket{\psi_{\rm in}}\bra{\psi_{\rm in}}$, $
\Pi_{\bar X}
=\mathbb I_X\otimes
\ket{\psi^{\rm in}_{\bar X}}\bra{\psi^{\rm in}_{\bar X}},
$
and $T=T_{(1,2)}$. The postselection projector $\Pi_{\bar X}$ acts on
the full $N$-qubit Hilbert space. Thus
$\Pi_{\bar X}\rho_{\rm in}=\rho_{\rm in}$,
$\mathrm{Tr}[\Pi_{\bar X}]=D=2^k$, and
$\mathrm{Tr}\rho_{\rm in}=\mathrm{Tr}\rho_{\rm in}^2=1$.
For each encoder, we define $n_U=\mathrm{Tr}(\rho_{\rm in}\rho_{\rm out})$, $p_U=\mathrm{Tr}\!\left(\Pi_{\bar X}\rho_{\rm out}\right)$, and the EED fidelity is $F = n_{U}/p_{U}$ when $p_U>0$; for $p_U=0$ we set $F=0$ in ensemble averages, since $n_U=0$.

With $\bar n_{\mathcal E}=\mathbb E n_U$ and
$\bar p_{\mathcal E}=\mathbb E p_U>0$,
the precise relation to the fidelity in Eq.~(\ref{fidelity}) is
\begin{equation}
\tilde F
=\frac{\bar n_{\mathcal E}}{\bar p_{\mathcal E}}
=\frac{\mathbb E[p_UF]}{\bar p_{\mathcal E}},
\qquad
\tilde F-\overline{F}
=\frac{\operatorname{Cov}(p_U,F)}
{\bar p_{\mathcal E}}.
\label{e:anneal}
\end{equation}
Thus $\tilde F$ is obtained by pooling accepted trials
with independently sampled encoders.

To exhibit the second moments explicitly, let
$A_K=\hat K\otimes\hat K^\dagger$, reserving $A$ for the
doping gate.
The identity $\mathrm{Tr}[(X\otimes Y)T]=\mathrm{Tr}(XY)$ gives
\begin{equation}
\begin{split}
\bar n_{\mathcal E}
&=\mathrm{Tr}\!\left[
\Phi_{\mathcal E}^{(2)}(A_K)B_{\rm num}^{F}
\right],\\
\bar p_{\mathcal E}
&=\mathrm{Tr}\!\left[
\Phi_{\mathcal E}^{(2)}(A_K)B_{\rm den}^{F}
\right],\\
B_{\rm num}^{F}
&=\rho_{\rm in}^{\otimes2},\\
B_{\rm den}^{F}
&=\left(\rho_{\rm in}\otimes\Pi_{\bar X}\right)T.
\end{split}
\label{e:contractions}
\end{equation}
We use linear vectorization with
$\langle\!\langle X|Y\rangle\!\rangle=\mathrm{Tr}(X^\dagger Y)$,
and denote the matrix of $\Phi_{\mathcal E}^{(2)}$ in this
operator space by $M_{\mathcal E}$.
The Hilbert--Schmidt norm of this matrix is its Frobenius
norm, not the induced norm of the channel.
For independent $U,V$,
\begin{equation}
\begin{split}
\|M_{\mathcal E}\|_{\rm HS}^{2}
&=\mathbb E_{U,V}|\mathrm{Tr}(U^\dagger V)|^4
=\mathcal F_{\mathcal E}^{(2)},\\
\|M_{\mathcal E}-M_{\rm Haar}\|_{\rm HS}^{2}
&=\mathcal F_{\mathcal E}^{(2)}-2=\Delta_{\mathcal E}.
\end{split}
\label{e:norm}
\end{equation}
Indeed, Haar invariance implies
$M_{\rm Haar}^2=M_{\rm Haar}=M_{\rm Haar}^\dagger$
and
$M_{\rm Haar}M_{\mathcal E}
=M_{\mathcal E}M_{\rm Haar}
=M_{\rm Haar}$.
The range of $M_{\rm Haar}$ is spanned by $\mathbb I$ and $T$, so its rank
is two.
Expanding the squared norm yields the second identity,
the standard relation between the frame potential and
the Hilbert--Schmidt distance of the second-moment
operators~\cite{Roberts:2017tk}.

Write $f=\tilde F_{\rm Haar}$,
$B_f=B_{\rm num}^{F}-fB_{\rm den}^{F}$, and
$\Delta\Phi=\Phi_{\mathcal E}^{(2)}
-\Phi_{\rm Haar}^{(2)}$.
Since the Haar contraction with $B_f$ vanishes,
Eq.~\eqref{e:contractions} gives the exact contrast
\begin{equation}
\bar p_{\mathcal E}(\tilde F-f)
=\mathrm{Tr}[\Delta\Phi(A_K)B_f].
\label{e:contrast}
\end{equation}
Unitarity of $\hat K$ gives $\|A_K\|_{\rm HS}=d$.
Purity of the input and
$\Pi_{\bar X}\rho_{\rm in}=\rho_{\rm in}$ give
\begin{equation}
\begin{split}
\|B_{\rm num}^{F}\|_{\rm HS}^2
&=1,
\qquad
\|B_{\rm den}^{F}\|_{\rm HS}^2=D,\\
\mathrm{Tr}[(B_{\rm num}^{F})^\dagger B_{\rm den}^{F}]
&=1,
\qquad
\|B_f\|_{\rm HS}^2=1+Df^2-2f.
\end{split}
\label{e:bnorm}
\end{equation}
Cauchy--Schwarz in operator space and the inequality
between induced and Frobenius norms now yield
\begin{equation}
\begin{split}
|\bar p_{\mathcal E}(\tilde F-f)|
&\leq
\|B_f\|_{\rm HS}\,
\|\Delta\Phi(A_K)\|_{\rm HS}\\
&\leq
d\sqrt{1+Df^2-2f}\,
\sqrt{\Delta_{\mathcal E}}.
\end{split}
\label{e:CS}
\end{equation}
For $k>0$,
$1+Df^2-2f=(1-f)^2+(D-1)f^2>0$.
Squaring and dividing proves Eq.~(\ref{eq:eed_witness}) for any encoder
distribution with $\bar p_{\mathcal E}>0$ and any fixed unitary error layer.

\subsection*{Shell sectors and the dark gap mode.}
The full doped-matchgate shell matrix $C$ acts on Majorana weights
$\ell=0,\ldots,2N$. Since $C$ only connects $\ell$ to $\ell$ and $\ell\pm2$, it splits into even and odd shells, $C=C_{\rm even}\oplus C_{\rm odd}$, and Eq.~(\ref{eq6}) only involves $\Gamma=C_{\rm even}$.
Its fixed space is four-dimensional and spanned by $\{\mathbb{I},T,\hat P\otimes\hat P,T(\hat P\otimes\hat P)\}$, with $T$ the replica swap and $\hat P$ the fermion parity.
In the shell basis we can take as fixed vectors the two endpoints $\ell=0$ and $\ell=2N$, an even-shell bulk vector with binomial weights, and an odd-shell bulk vector, i.e., three even and one odd at $\lambda=1$.

The first \emph{decaying} quartet has the opposite allocation. The four
invariant birth--death sectors of the full transfer problem share the
Ornstein--Uhlenbeck (OU) levels
$\lambda_j=1-2j/N+O(N^{-2})$~\cite{dopeddesigns}. Resolving the $j=1$ level
in shell parity gives one even-shell mode and three odd-shell modes, all
approaching $1-2/N$. These are first excited modes, rather than members of
the fixed space. The frame potential retains all four and is quadratic in
their amplitudes. Hence, with $\tau=n_d/N$ and $\Phi_\infty^{(2)}$ the
second-moment channel of the symmetry-resolved Haar class
$\mathcal E_\infty$ reached as $n_d\to\infty$,
\begin{equation}
 \begin{split}
 \lambda_{\rm gap}&=1-\frac{2}{N}+O(N^{-3}),\\
 \epsilon_{\rm HS}^{2}&\equiv
 \|M_{\rm D}-M_{\infty}\|_{\rm HS}^2
 =\mathcal F_{\rm D}^{(2)}-\mathcal F_{\infty}^{(2)}
 \simeq4e^{-4\tau},
 \end{split}
 \label{eq:endmatter-frame}
\end{equation}
and hence $\epsilon_{\rm HS}\simeq2e^{-2\tau}$. In particular, the rate
$4$ of the frame-potential excess comes from
$\lambda_{\rm gap}^{2n_d}\to e^{-4\tau}$.

The error vector and both boundaries of the EED fidelity have only even-shell support. To classify the remaining mode, define shell reflection by
$(Rv)_m=v_{N-m}$ and introduce
$S=\operatorname{diag}[(-1)^m]_{m=0}^{N}$.
In the positive staggered-sign gauge,
$\Gamma_+=S\Gamma S$ commutes with $R$.
Its stationary bulk vector is positive and $R$-even,
whereas its first decaying bulk eigenvector
$w_{\rm gap}=Sv_{\rm gap}$ has a single node and is
$R$-odd; in the continuum limit it becomes the odd
OU eigenfunction $H_1(x)\propto x$.
For the EED boundary vectors
$b_{\rm num}=(N_0,N_2,\ldots,N_{2N})^T$ and
$b_{\rm den}=(D_0,D_2,\ldots,D_{2N})^T$,
the transformed vectors $Sb_a$ are $R$-even.
Hence
\begin{equation}
b_a^Tv_{\rm gap}
=(Sb_a)^Tw_{\rm gap}
=(RSb_a)^T(Rw_{\rm gap})
=-b_a^Tv_{\rm gap}=0,
\label{eq:endmatter-dark}
\end{equation}
where $a=\mathrm{num},\mathrm{den}$.
This removes the entire first decaying level from the fidelity.

\subsection*{Fidelity response and design accuracy.}
To determine which modes are seen by the fidelity, write
$h^{(0)}=(s_0,s_2,\ldots,s_{2N})^T$ for the pre-doping shell vector and
$h_{n_d}=\Gamma^{n_d}h^{(0)}=h_{\infty}+\delta h$. Define
$p_a=b_a^Th_{n_d}=p_{a,\infty}+\delta p_a$ for
$a=\mathrm{num},\mathrm{den}$.
These shell contractions satisfy
$\mathbb E[n_U]=d\,p_{\rm num}$ and
$\mathbb E[p_U]=d\,p_{\rm den}$, so that
$\tilde F_{\mathrm D}=p_{\rm num}/p_{\rm den}$.
Here the averages are over the doped-matchgate ensemble.
Expanding $\tilde F_{\mathrm D}=p_{\rm num}/p_{\rm den}$
about $\tilde F_\infty=p_{{\rm num},\infty}/p_{{\rm den},\infty}$
gives
\begin{align}
\delta\tilde F_{\mathrm D}
&\equiv \tilde F_{\mathrm D}-\tilde F_\infty
=\frac{\delta p_{\rm num}
-\tilde F_\infty\delta p_{\rm den}}
{p_{{\rm den},\infty}}
+O(\|\delta h\|^2),
\label{eq:endmatter-response}\\
\delta p_a
&=b_a^T\delta h
=\sum_j c_j\lambda_j^{n_d}b_a^Tv_j.
\label{eq:endmatter-modes}
\end{align}
Here the sum runs over decaying modes of $\Gamma$, with orthonormal
eigenvectors $v_j$ and $c_j=v_j^Th^{(0)}$.
A mode thus contributes to the linear response only if
it is excited by the input, $c_j\neq0$, and has a nonzero
overlap with $b_{\rm num}-\tilde F_\infty b_{\rm den}$.

Because the two selection rules above remove the $j=1$ quartet, the first
boundary-visible contribution is the reflection-even $j=2$, or bright,
mode,
\begin{align}
 \lambda_{\rm bright}&=1-\frac{4}{N}+O(N^{-2}),\nonumber\\
 \tilde F_{\infty}-\tilde F_{\mathrm D}
 &\simeq Ng(\alpha)\lambda_{\rm bright}^{n_d}
 \sim Ng(\alpha)e^{-4\tau}.
 \label{eq:endmatter-bright}
\end{align}
This explains the observed slopes: the frame potential squares the
level-one decay, while the fidelity is linear in the first visible
level-two decay. Inside the protected window, $\tilde F_\infty$ and $\tilde F_{\rm Haar}$ agree up to corrections that are exponentially small in $N$.

Combining Eqs.~\eqref{eq:endmatter-frame} and \eqref{eq:endmatter-bright} at late doping, we obtain
\begin{equation}
 \tilde F_{\infty}-\tilde F_{\mathrm D}
 \simeq \frac{Ng(\alpha)}{4}\epsilon_{\rm HS}^2 .
 \label{eq:endmatter-hsrelation}
\end{equation}
The protected-window average inherits the scaling $Ne^{-4\tau}$. To keep this deviation fixed in the scaling limit, one must fix $\epsilon_{\rm HS}=O(N^{-1/2})$, implying that we need $n_d=\Theta(N\log N)$, consisten with the design lower bound in~\cite{dopeddesigns}.

The required scaling of an approximate-design error depends on its
definition. Here $\epsilon_{\rm HS}$ is an additive error, read off directly from the frame potential. If the numerator and denominator of the EED fidelity, both positive, are known up to a relative error $\epsilon_{\rm rel}$, the fidelity is off by at most $2\epsilon_{\rm rel}/(1-\epsilon_{\rm rel})$. An additive bound, for instance in diamond norm, has to be small compared with the exponentially small postselection probability, and can therefore require a much higher accuracy.
This estimate breaks down close to $\alpha_c$, where the crossing point is very sensitive to the doping. Pinning it down to within $\delta_N$ takes roughly $(N/4)\log(1/\delta_N)$ extra gates. 

\bibliography{reference_enc}

\end{document}


\title{Supplemental Material for\\
``Error Resilience and Unitary 2-Designs''}
\author{Zheng-Hang Sun, Fabian Ballar Trigueros, Xhek Turkeshi, and Piotr Sierant}
\maketitle

\section{Derivation of Eqs.~(4)--(6) of the main text}
\label{sec:matchgate_fidelity_derivation}

In this section, we derive the annealed fidelity in Eq.~(5) of the main text and the two boundary coefficients $N_{2m}$ and $D_{2m}$ in Eq.~(6).  We use precisely the conventions of the main text: the system contains $N$ qubits, $X$ contains $k$ logical qubits, $\bar X$ contains $N-k$ ancilla qubits, and the matchgate calculation uses
\begin{equation}
 \rho_{\mathrm{in}}=\ket{\psi_{\mathrm{in}}}\!\bra{\psi_{\mathrm{in}}}=
 \ket{0}\!\bra{0}^{\otimes N},
 \qquad
 \rho_{\bar X}^{\mathrm{in}}=\ket{0_{\bar X}}\!\bra{0_{\bar X}},
 \qquad
 \Pi_{\bar X}=\id_X\otimes\rho_{\bar X}^{\mathrm{in}}.
 \label{eq:SM_initial_data}
\end{equation}
The local coherent error is
\begin{equation}
 \hat K=\exp\!\left(-\frac{i\alpha}{2}\sum_{j=1}^{N}\sigma_j^z\right),
 \label{eq:SM_K_def}
\end{equation}
and, for a matchgate encoder $U_{\mathrm M}$, we write
\begin{equation}
 V_{U_{\mathrm M}}=U_{\mathrm M}^\dagger \hat K U_{\mathrm M}.
 \label{eq:SM_V_def}
\end{equation}

\subsection{Replica representation of the annealed fidelity}

For a fixed realization of $U_{\mathrm M}$, the unnormalized numerator and postselection probability entering the fidelity are
\begin{align}
 n(U_{\mathrm M})
 &=\Tr\!\left(\rho_{\mathrm{in}} V_{U_{\mathrm M}}\rho_{\mathrm{in}} V_{U_{\mathrm M}}^\dagger\right),
 \label{eq:SM_fixed_numerator}\\
 p(U_{\mathrm M})
 &=\Tr\!\left(\Pi_{\bar X} V_{U_{\mathrm M}}\rho_{\mathrm{in}} V_{U_{\mathrm M}}^\dagger\right).
 \label{eq:SM_fixed_denominator}
\end{align}
Let $T_{(1,2)}$ swap the two replicas.  The identity
\begin{equation}
 \Tr\!\left[(A\otimes B)(C\otimes D)T_{(1,2)}\right]
 =\Tr(ACBD)
 \label{eq:SM_swap_identity}
\end{equation}
allows Eqs.~\eqref{eq:SM_fixed_numerator} and \eqref{eq:SM_fixed_denominator} to be written as
\begin{align}
 n(U_{\mathrm M})
 &=\Tr\!\left[(V_{U_{\mathrm M}}\otimes V_{U_{\mathrm M}}^\dagger)B_{\mathrm{num}}^F\right],
 &B_{\mathrm{num}}^F&=\rho_{\mathrm{in}}^{\otimes 2},
 \label{eq:SM_Bnum}\\
 p(U_{\mathrm M})
 &=\Tr\!\left[(V_{U_{\mathrm M}}\otimes V_{U_{\mathrm M}}^\dagger)B_{\mathrm{den}}^F\right],
 &B_{\mathrm{den}}^F&=(\rho_{\mathrm{in}}\otimes \Pi_{\bar X})T_{(1,2)}.
 \label{eq:SM_Bden}
\end{align}
The second-moment matchgate twirl defined in the main text obeys
\begin{equation}
 \mathbb E_{U_{\mathrm M}}\!\left[V_{U_{\mathrm M}}\otimes V_{U_{\mathrm M}}^\dagger\right]
 =\Phi_{\mathrm M}^{(2)}(\hat K\otimes\hat K^\dagger).
 \label{eq:SM_twirl_average}
\end{equation}
Taking the ratio of the separately averaged quantities gives the annealed fidelity,
\begin{equation}
 \boxed{
 \tilde F_{\mathrm M}
 =\frac{\mathbb E_{U_{\mathrm M}}[n(U_{\mathrm M})]}{\mathbb E_{U_{\mathrm M}}[p(U_{\mathrm M})]}
 =\frac{\Tr\!\left[\Phi_{\mathrm M}^{(2)}(\hat K\otimes\hat K^\dagger)B_{\mathrm{num}}^F\right]}
 {\Tr\!\left[\Phi_{\mathrm M}^{(2)}(\hat K\otimes\hat K^\dagger)B_{\mathrm{den}}^F\right]} .}
 \label{eq:SM_annealed_fidelity}
\end{equation}
This is Eq.~(2) of the main text specialized to the matchgate ensemble.  Notice that $\tilde F_{\mathrm M}$ is a ratio of ensemble averages, whereas $\mathbb E_{U_{\mathrm M}}[F]$ is the average of the sample-dependent ratio $n(U_{\mathrm M})/p(U_{\mathrm M})$.  Their numerical agreement is examined separately in the Supplemental Material.

\subsection{Majorana-shell decomposition of the error operator}

Let $d=2^N$ and introduce $2N$ Majorana operators satisfying $\{\gamma_\mu,\gamma_\nu\}=2\delta_{\mu\nu}$.  We use
\begin{equation}
 \sigma_j^z=-i\gamma_{2j-1}\gamma_{2j},
 \qquad
 \gamma_S=\prod_{\mu\in S}^{\nearrow}\gamma_\mu,
 \qquad
 \ketket{\gamma_S/\sqrt d}=\frac{\operatorname{vec}(\gamma_S)}{\sqrt d},
 \label{eq:SM_majorana_basis}
\end{equation}
where the arrow fixes increasing Majorana order.  The basis is orthonormal under the Hilbert--Schmidt inner product.  For the full orthogonal Gaussian ensemble used in the main text, the two-copy invariant shell vectors are
\begin{equation}
 \ketket{\Upsilon_\ell}
 =\binom{2N}{\ell}^{-1/2}
 \sum_{\substack{S\subseteq[2N]\\|S|=\ell}}
 \ketket{\gamma_S/\sqrt d}\otimes\ketket{\gamma_S/\sqrt d},
 \qquad \ell=0,\ldots,2N,
 \label{eq:SM_Upsilon_def}
\end{equation}
Here $\ketket{O}=\operatorname{vec}(O)$ denotes linear,
tensor-compatible vectorization, with
$\braketket{O}{P}=\Tr(O^\dagger P)$.
Thus $\ketket{\Upsilon_\ell}=\operatorname{vec}(\Upsilon_\ell)$,
where $\Upsilon_\ell$ is the normalized two-copy operator in
Eq.~(4) of the main text.  The second-moment twirl is the orthogonal
projector onto their span.

Because $(\gamma_{2j-1}\gamma_{2j})^2=-\id$, Eq.~\eqref{eq:SM_K_def} factorizes as
\begin{align}
 \hat K
 &=\prod_{j=1}^{N}\left[
 \cos\!\left(\frac\alpha2\right)\id
 -\sin\!\left(\frac\alpha2\right)\gamma_{2j-1}\gamma_{2j}
 \right],
 \label{eq:SM_K_expand}\\
 \hat K^\dagger
 &=\prod_{j=1}^{N}\left[
 \cos\!\left(\frac\alpha2\right)\id
 +\sin\!\left(\frac\alpha2\right)\gamma_{2j-1}\gamma_{2j}
 \right].
 \label{eq:SM_Kdag_expand}
\end{align}
Only monomials made of complete canonical pairs $\{2j-1,2j\}$ occur.  Consequently, only even Majorana shells contribute.  Projecting the two-copy operator onto Eq.~\eqref{eq:SM_Upsilon_def} gives
\begin{equation}
 \operatorname{vec}\!\left[
 \Phi_{\mathrm M}^{(2)}(\hat K\otimes\hat K^\dagger)\right]
 =d\sum_{m=0}^{N}s_{2m}\ketket{\Upsilon_{2m}},
 \label{eq:SM_twirl_shell_expansion}
\end{equation}
in agreement with the operator normalization of the main text.  There are $\binom Nm$ choices of $m$ canonical pairs, and the product of the coefficients from Eqs.~\eqref{eq:SM_K_expand} and \eqref{eq:SM_Kdag_expand} is $(-1)^m\sin^{2m}(\alpha/2)\cos^{2(N-m)}(\alpha/2)$.  Therefore
\begin{equation}
 s_{2m}
 =\binom{2N}{2m}^{-1/2}\binom Nm(-1)^m
 \left(\cos^2\frac\alpha2\right)^{N-m}
 \left(\sin^2\frac\alpha2\right)^m,
 \qquad
 s_{2m+1}=0,
 \label{eq:SM_s2m}
\end{equation}
which reproduces $s_{2m}$ in Eq.~(6) of the main text.

\subsection{Numerator boundary and Eq.~(6)}

Define the shell overlap with the numerator boundary by
\begin{equation}
 N_{2m}=\braketket{\Upsilon_{2m}}{B_{\mathrm{num}}^F}.
 \label{eq:SM_N_def}
\end{equation}
Using Eqs.~\eqref{eq:SM_Bnum} and \eqref{eq:SM_Upsilon_def},
\begin{equation}
 N_{2m}
 =\binom{2N}{2m}^{-1/2}\frac1d
 \sum_{|S|=2m}
 \Tr(\gamma_S^\dagger\rho_{\mathrm{in}})\,
 \Tr(\gamma_S^\dagger\rho_{\mathrm{in}}).
 \label{eq:SM_N_sum}
\end{equation}
The vacuum expectation value vanishes if $S$ contains exactly one Majorana from any physical site.  Hence, a nonzero contribution requires
\begin{equation}
 S=\bigcup_{j\in J}\{2j-1,2j\},
 \qquad |J|=m,
 \label{eq:SM_canonical_S}
\end{equation}
and there are $\binom Nm$ such subsets.  Since
\begin{equation}
 \gamma_S=i^m\prod_{j\in J}\sigma_j^z,
 \qquad
 \bra{0^{\otimes N}}\gamma_S^\dagger\ket{0^{\otimes N}}=(-i)^m,
 \label{eq:SM_vacuum_phase}
\end{equation}
the product of the two identical vacuum matrix elements in Eq.~\eqref{eq:SM_N_sum} is $(-1)^m$.  We obtain
\begin{equation}
 \boxed{
 N_{2m}
 =2^{-N}\binom{2N}{2m}^{-1/2}\binom Nm(-1)^m .}
 \label{eq:SM_N_result}
\end{equation}
This is $N_{2m}$ in Eq.~(6) of the main text.

\subsection{Denominator boundary and Eq.~(6)}

We now evaluate
\begin{equation}
 D_{2m}=\braketket{\Upsilon_{2m}}{B_{\mathrm{den}}^F}.
 \label{eq:SM_D_def}
\end{equation}
For each $S$, Eqs.~\eqref{eq:SM_swap_identity} and \eqref{eq:SM_Bden} give the elementary boundary matrix element
\begin{align}
 b_S^{(\mathrm{den})}
 &\equiv \frac1d
 \Tr\!\left[(\gamma_S^\dagger\otimes\gamma_S^\dagger)
 (\rho_{\mathrm{in}}\otimes \Pi_{\bar X})T_{(1,2)}\right]
 \nonumber\\
 &=\frac1d\Tr(\gamma_S^\dagger\rho_{\mathrm{in}}\gamma_S^\dagger \Pi_{\bar X})
 =\frac1d\bra{0^{\otimes N}}\gamma_S^\dagger \Pi_{\bar X}
 \gamma_S^\dagger\ket{0^{\otimes N}}.
 \label{eq:SM_elementary_D}
\end{align}
Acting on the computational-basis vacuum, a Majorana monomial flips precisely those physical qubits on which it contains an odd number of the two local Majoranas.  Since $\Pi_{\bar X}$ projects the ancillas in $\bar X$ back onto their vacuum while acting as the identity on $X$, Eq.~\eqref{eq:SM_elementary_D} is nonzero if and only if every site in $\bar X$ contributes either zero or two Majoranas.  No analogous restriction occurs on $X$.

Suppose $m_{\bar X}$ sites in $\bar X$ contribute complete Majorana pairs.  They account for $2m_{\bar X}$ of the total $2m$ Majoranas and can be chosen in $\binom{N-k}{m_{\bar X}}$ ways.  The remaining $2(m-m_{\bar X})$ Majoranas can be chosen arbitrarily among the $2k$ Majoranas supported in $X$.  The total number of allowed subsets is thus
\begin{equation}
 A_{2m}(k,N-k)
 =\sum_{m_{\bar X}=0}^{\min(m,N-k)}
 \binom{N-k}{m_{\bar X}}
 \binom{2k}{2(m-m_{\bar X})}.
 \label{eq:SM_allowed_count}
\end{equation}
Here and below, $\binom ab=0$ when $b<0$ or $b>a$, so Eq.~\eqref{eq:SM_allowed_count} automatically enforces $2(m-m_{\bar X})\leq 2k$.

For every allowed $S$, the projector $\Pi_{\bar X}$ acts trivially on $\gamma_S^\dagger\ket{0^{\otimes N}}$.  Moreover, for $|S|=2m$,
\begin{equation}
 (\gamma_S^\dagger)^2=\gamma_S^2
 =(-1)^{|S|(|S|-1)/2}\id=(-1)^m\id.
 \label{eq:SM_majorana_square}
\end{equation}
Therefore every allowed subset contributes the same value $b_S^{(\mathrm{den})}=(-1)^m2^{-N}$. 
\begin{equation}
 b_S^{(\mathrm{den})}=(-1)^m2^{-N}.
 \label{eq:SM_allowed_value}
\end{equation}
Substituting Eqs.~\eqref{eq:SM_allowed_count} and \eqref{eq:SM_allowed_value} into Eq.~\eqref{eq:SM_D_def}, we arrive at
\begin{equation}
 \boxed{
 D_{2m}
 =(-1)^m2^{-N}\binom{2N}{2m}^{-1/2}
 \sum_{m_{\bar X}=0}^{\min(m,N-k)}
 \binom{N-k}{m_{\bar X}}
 \binom{2k}{2(m-m_{\bar X})}.}
 \label{eq:SM_D_result}
\end{equation}
This is Eq.~(6) of the main text.

\subsection{Final form and normalization check}

With the normalized Majorana basis of Eq.~\eqref{eq:SM_majorana_basis}, the two contractions entering Eq.~\eqref{eq:SM_annealed_fidelity} are
\begin{align}
 \mathbb E_{U_{\mathrm M}}[n(U_{\mathrm M})]
 &=d\sum_{m=0}^{N}s_{2m}N_{2m},
 \label{eq:SM_final_num}\\
 \mathbb E_{U_{\mathrm M}}[p(U_{\mathrm M})]
 &=d\sum_{m=0}^{N}s_{2m}D_{2m}.
 \label{eq:SM_final_den}
\end{align}
The common factor $d=2^N$ cancels in their ratio.  Equivalently, defining $A_{2m}$ by Eq.~\eqref{eq:SM_allowed_count}, the annealed fidelity can be written without any normalization ambiguity as
\begin{equation}
 \tilde F_{\mathrm M}
 =\frac{
 \displaystyle\sum_{m=0}^{N}
 \frac{\binom Nm^2}{\binom{2N}{2m}}
 \left(\sin^2\frac\alpha2\right)^m
 \left(\cos^2\frac\alpha2\right)^{N-m}}
 {\displaystyle\sum_{m=0}^{N}
 \frac{\binom Nm A_{2m}(k,N-k)}{\binom{2N}{2m}}
 \left(\sin^2\frac\alpha2\right)^m
 \left(\cos^2\frac\alpha2\right)^{N-m}}.
 \label{eq:SM_final_ratio}
\end{equation}
The factors $(-1)^m$ in $s_{2m}$ cancel those in both boundary coefficients, so every term in Eq.~\eqref{eq:SM_final_ratio} is nonnegative.  Several limiting cases provide useful checks.  At $\alpha=0$, only $m=0$ contributes and $\tilde F_{\mathrm M}=1$.  If $k=0$, then $A_{2m}=\binom Nm$ and the numerator equals the denominator, again giving $\tilde F_{\mathrm M}=1$.  If $N-k=0$, then $A_{2m}=\binom{2N}{2m}$ and Eq.~\eqref{eq:SM_final_den} gives unit postselection probability.

\subsection{Thermodynamic limit at half code rate}
\label{subsec:SM_matchgate_thermodynamic_limit}

We now prove the thermodynamic-limit statement used in the main
text.  Let $N$ be even and set
$k=N/2$.  It is convenient to introduce $x=\sin^2\!\left(\frac{\alpha}{2}\right)$, $y=\cos^2\!\left(\frac{\alpha}{2}\right)$, $x+y=1$, 
and the nonnegative coefficients
\begin{equation}
 u_m=\frac{\binom Nm^2}{\binom{2N}{2m}},
 \qquad
 v_m=\frac{\binom Nm A_{2m}(N/2,N/2)}
 {\binom{2N}{2m}}.
 \label{eq:SM_uv_def}
\end{equation}
Equation~\eqref{eq:SM_final_ratio} then becomes
\begin{equation}
 \tilde F_{\mathrm M,N}(\alpha)
 =\frac{\displaystyle\sum_{m=0}^{N}u_m x^m y^{N-m}}
 {\displaystyle\sum_{m=0}^{N}v_m x^m y^{N-m}}.
 \label{eq:SM_F_uv}
\end{equation}

First consider $0<x<y$.  Vandermonde's identity gives
\begin{equation}
 \binom{2N}{2m}
 =\sum_j\binom Nj\binom N{2m-j}
 \geq \binom Nm^2,
 \label{eq:SM_vandermonde_bound}
\end{equation}
and hence $0\leq u_m\leq1$.  With $z=x/y<1$, the numerator
of Eq.~\eqref{eq:SM_F_uv} therefore satisfies the uniform bound
\begin{equation}
 \sum_{m=0}^{N}u_m x^m y^{N-m}
 =y^N\sum_{m=0}^{N}u_m z^m
 \leq \frac{y^N}{1-z}.
 \label{eq:SM_num_thermo_bound}
\end{equation}
All denominator terms are nonnegative, so it is sufficient to retain
only the $m=1$ term.  At half code rate,
\begin{equation}
 A_2(N/2,N/2)
 =\binom{N/2}{1}+\binom N2=\frac{N^2}{2},
 \qquad
 v_1=\frac{N^2}{2(2N-1)}.
 \label{eq:SM_v1_exact}
\end{equation}
Combining Eqs.~\eqref{eq:SM_num_thermo_bound} and
\eqref{eq:SM_v1_exact}, we obtain the explicit upper bound
\begin{equation}
 0\leq \tilde F_{\mathrm M,N}(\alpha)
 \leq
 \frac{2(2N-1)y^2}{N^2x(y-x)}
 =O\!\left(\frac1N\right),
 \qquad 0<x<y,
 \label{eq:SM_F_thermo_bound}
\end{equation}
where the constant implicit in $O(N^{-1})$ is finite for every fixed
$x\in(0,1/2)$.  Consequently,
$\lim_{N\to\infty}\tilde F_{\mathrm M,N}(\alpha)=0$ throughout this
interval.

The midpoint $x=y=1/2$ follows separately.  The numerator is at most
$(N+1)2^{-N}$.  In the denominator, the $m=2$ term alone gives
\begin{equation}
 \sum_{m=0}^{N}v_m2^{-N}
 \geq 2^{-N}\binom N2
 \frac{A_4(N/2,N/2)}{\binom{2N}{4}}
 \geq 2^{-N}\binom N2
 \frac{\binom N4}{\binom{2N}{4}}.
 \label{eq:SM_midpoint_den_bound}
\end{equation}
The last inequality keeps only the allowed subsets whose four
Majoranas all lie in $X$.  Since
$\binom N4/\binom{2N}{4}\to1/16$, it is at least $1/32$ for all
sufficiently large $N$.  Thus
\begin{equation}
 0\leq \tilde F_{\mathrm M,N}\!\left(\frac{\pi}{2}\right)
 \leq \frac{32(N+1)}{\binom N2}
 =O\!\left(\frac1N\right)
 \xrightarrow[N\to\infty]{}0.
 \label{eq:SM_midpoint_F_bound}
\end{equation}

It remains to note an exact reflection symmetry.  Taking the
complement of an allowed $2m$-Majorana subset gives an allowed
$2(N-m)$-Majorana subset: on every site in $\bar X$, the local
occupation changes from zero to two or from two to zero.  Therefore
\begin{equation}
 A_{2(N-m)}(N/2,N/2)=A_{2m}(N/2,N/2),
 \qquad u_{N-m}=u_m,
 \qquad v_{N-m}=v_m.
 \label{eq:SM_shell_reflection}
\end{equation}
Equation~\eqref{eq:SM_F_uv} is consequently invariant under
$x\leftrightarrow y$, which extends the preceding result to
$0<y<x$.  We have proved
\begin{equation}
 \boxed{
 \lim_{N\to\infty}\tilde F_{\mathrm M,N}(\alpha)=0
 \quad\text{for every fixed }\alpha\notin\pi\mathbb Z,
 \qquad k=N/2.}
 \label{eq:SM_matchgate_thermodynamic_result}
\end{equation}
Equivalently, because the exact finite-$N$ expression is invariant
under $x\leftrightarrow y$, one may use the fundamental interval
$0\leq\alpha\leq\pi/2$; within the interval studied in the main
text, every fixed $\alpha>0$ has vanishing thermodynamic fidelity.
The isolated angles $\alpha\in\pi\mathbb Z$ are trivial exceptions:
only $m=0$ or $m=N$ contributes to
Eq.~\eqref{eq:SM_final_ratio}, and $\tilde F_{\mathrm M,N}=1$ exactly.  The
bound is a fixed-$\alpha$ statement and is not uniform as
$\alpha\to0$, consistently allowing the finite-size
error-protecting window to shrink toward the origin as $N$ grows.

\section{Derivation of Eq.~(7) of the main text}
\label{sec:SM_doped_matchgate_derivation}

We now extend the preceding matchgate calculation to the doped
encoder in Fig.~1(b) of the main text,
\begin{equation}
 U_{\mathrm D}
 =U_{\mathrm M}^{(n_d)}A U_{\mathrm M}^{(n_d-1)}A\cdots A U_{\mathrm M}^{(0)},
 \qquad
 A=\exp\!\left(-\frac{i\pi}{4}Z_1Z_2\right),
 \label{eq:SM_doped_circuit}
\end{equation}
where the $U_{\mathrm M}^{(i)}$ are independent random matchgate circuits.
The derivation uses two results established in
Ref.~\cite{BallarTrigueros2026}.  First, the matchgate two-copy twirl
is the orthogonal projector onto the commutant spanned by the pairing
vectors $\ketket{\Upsilon_\ell}$ defined in
Eq.~\eqref{eq:SM_Upsilon_def}.  Second, after projection onto this
commutant, successive globally scrambled doping rounds are described
by repeated application of a finite-dimensional transfer matrix.  In
the notation of Ref.~\cite{BallarTrigueros2026}, the iterative channel
is given in its Eq.~(22), while the pairing basis is defined in its
Eq.~(27).  That reference uses $n$ for the
number of qubits and $2n$ for the number of Majoranas; here we retain
the main-text convention of $N$ qubits and $2N$ Majoranas.

\subsection{Transfer-matrix form of the doped twirl and Eq.~(7)}

For the Heisenberg-picture encoding-error-decoding (EED) echo, define the two-copy doping map
\begin{equation}
 \mathcal A_{\mathrm H}^{(2)}(O_1\otimes O_2)
 =(A^\dagger O_1A)\otimes(A^\dagger O_2A).
 \label{eq:SM_AH_def}
\end{equation}
Independence of all matchgate layers in
Eq.~\eqref{eq:SM_doped_circuit}, together with the idempotence of the
matchgate twirl, gives
\begin{align}
 &\mathbb E_{U_{\mathrm D}}\!\left[
 (U_{\mathrm D}^\dagger\hat K U_{\mathrm D})\otimes
 (U_{\mathrm D}^\dagger\hat K^\dagger U_{\mathrm D})\right]
 \nonumber\\
 &\hspace{9mm}=
 \left[\Phi_{\mathrm M}^{(2)}\circ\mathcal A_{\mathrm H}^{(2)}
 \circ\Phi_{\mathrm M}^{(2)}\right]^{n_d}
 \Phi_{\mathrm M}^{(2)}(\hat K\otimes\hat K^\dagger).
 \label{eq:SM_doped_twirl_iteration}
\end{align}
Reference~\cite{BallarTrigueros2026} writes the doping conjugation in
the Schr\"odinger direction, with $A$ and $A^\dagger$ interchanged
relative to Eq.~\eqref{eq:SM_AH_def}.  For the quartic gate in
Eq.~\eqref{eq:SM_doped_circuit}, the transfer matrix derived below is
real and symmetric and is unchanged under $A\leftrightarrow A^\dagger$.
Thus the same transfer matrix applies to the EED convention.

Within the orthonormal pairing basis, define
\begin{equation}
 C_{\ell',\ell}
 \equiv
 \left\langle\!\left\langle\Upsilon_{\ell'}\right|
 \mathcal A_{\mathrm H}^{(2)}
 \left|\Upsilon_{\ell}\right\rangle\!\right\rangle,
 \qquad \ell,\ell'=0,\ldots,2N.
 \label{eq:SM_C_def}
\end{equation}
The matchgate twirls surrounding
$\mathcal A_{\mathrm H}^{(2)}$ in
Eq.~\eqref{eq:SM_doped_twirl_iteration} act as the identity on these
commutant vectors.  Moreover, Eq.~\eqref{eq:SM_s2m} shows that the
initial error vector has support only on even Majorana shells.  Since
the quartic dopant preserves shell parity, it is therefore sufficient
to restrict $C$ to its even-shell block,
\begin{equation}
 \Gamma_{r,m}\equiv C_{2r,2m},
 \qquad r,m=0,\ldots,N.
 \label{eq:SM_Gamma_def}
\end{equation}
Using the shell expansion in
Eq.~\eqref{eq:SM_twirl_shell_expansion}, define
$h^{(0)}=(s_0,s_2,\ldots,s_{2N})^T$ and
$h_{n_d}=\Gamma^{n_d}h^{(0)}$, as in the End Matter.
Its component associated with shell $2r$ is consequently
\begin{equation}
 (h_{n_d})_r
 =\sum_{m=0}^{N}(\Gamma^{n_d})_{r,m}s_{2m}.
 \label{eq:SM_doped_shell_coefficients}
\end{equation}

The numerator and denominator boundaries are unchanged by the
doping.  Contracting Eq.~\eqref{eq:SM_doped_shell_coefficients} with
$N_{2r}$ and $D_{2r}$ from
Eqs.~\eqref{eq:SM_N_result} and \eqref{eq:SM_D_result}, respectively,
and cancelling the common factor $d$ identified
in Eqs.~\eqref{eq:SM_final_num} and \eqref{eq:SM_final_den}, yields
\begin{equation}
 \boxed{
 \tilde F_{\mathrm D}
 =\frac{\displaystyle
 \sum_{r,m=0}^{N}N_{2r}
 (\Gamma^{n_d})_{r,m}s_{2m}}
 {\displaystyle
 \sum_{r,m=0}^{N}D_{2r}
 (\Gamma^{n_d})_{r,m}s_{2m}}.}
 \label{eq:SM_doped_fidelity}
\end{equation}
This is Eq.~(7) of the main text.  At $n_d=0$,
$\Gamma^{0}=\id$, and Eq.~\eqref{eq:SM_doped_fidelity} reduces
immediately to the undoped result derived in the preceding section.

\subsection{Quartic-dopant matrix elements}

It remains to evaluate the transfer matrix in
Eq.~\eqref{eq:SM_C_def}.  Let
\begin{equation}
 \mathcal T=\{1,2,3,4\},
 \qquad
 G=\gamma_1\gamma_2\gamma_3\gamma_4=-Z_1Z_2,
 \qquad
 A=e^{i\pi G/4}.
 \label{eq:SM_active_majoranas}
\end{equation}
For a Majorana subset $S$, set $q=|S\cap \mathcal T|$.  Since
$G\gamma_S=(-1)^q\gamma_SG$, conjugation by $A$ gives
\begin{equation}
 A^\dagger\gamma_S A
 =
 \begin{cases}
 \gamma_S, & q=0,2,4,\\[1mm]
 -iG\gamma_S, & q=1,3.
 \end{cases}
 \label{eq:SM_A_action_single_copy}
\end{equation}
Up to the sign required to restore increasing Majorana order,
$G\gamma_S$ is the monomial associated with the symmetric difference
$S\mathbin{\triangle}\mathcal T$.  This reordering sign is squared in the
two-copy pairing vector and hence cancels.  Equation
\eqref{eq:SM_A_action_single_copy} therefore implies
\begin{equation}
 \mathcal A_{\mathrm H}^{(2)}
 \left(\ketket{\gamma_S/\sqrt d}\otimes\ketket{\gamma_S/\sqrt d}\right)
 =
 \begin{cases}
 \ketket{\gamma_S/\sqrt d}\otimes\ketket{\gamma_S/\sqrt d}, & q=0,2,4,\\[1mm]
 -\ketket{\gamma_{S\triangle \mathcal T}/\sqrt d}\otimes
  \ketket{\gamma_{S\triangle \mathcal T}/\sqrt d}, & q=1,3.
 \end{cases}
 \label{eq:SM_A_action_two_copy}
\end{equation}
Changing the sign of the exponent of $A$ changes $-i$ to $+i$ in
Eq.~\eqref{eq:SM_A_action_single_copy}, but leaves the minus sign in
Eq.~\eqref{eq:SM_A_action_two_copy} unchanged.  This explicitly
verifies the $A\leftrightarrow A^\dagger$ statement made above.

Consider an input shell of Majorana weight $\ell$.  The number of
subsets with $|S\cap \mathcal T|=q$ is
\begin{equation}
 \binom{4}{q}\binom{2N-4}{\ell-q}.
 \label{eq:SM_active_overlap_count}
\end{equation}
For even $q$, the monomial is unchanged.  Summing the contributions
with $q=0,2,4$ and using the normalization of
Eq.~\eqref{eq:SM_Upsilon_def} gives
\begin{equation}
 C_{\ell,\ell}
 =\frac{
 \binom{2N-4}{\ell}
 +6\binom{2N-4}{\ell-2}
 +\binom{2N-4}{\ell-4}}
 {\binom{2N}{\ell}}.
 \label{eq:SM_C_diagonal}
\end{equation}
For $q=1$, the symmetric difference has weight $\ell+2$; for
$q=3$, it has weight $\ell-2$.  The corresponding matrix elements
are
\begin{align}
 C_{\ell+2,\ell}
 &=-\frac{4\binom{2N-4}{\ell-1}}
 {\sqrt{\binom{2N}{\ell+2}\binom{2N}{\ell}}},
 \label{eq:SM_C_up}\\
 C_{\ell-2,\ell}
 &=-\frac{4\binom{2N-4}{\ell-3}}
 {\sqrt{\binom{2N}{\ell-2}\binom{2N}{\ell}}}.
 \label{eq:SM_C_down}
\end{align}
All other entries vanish.  The map
$S\mapsto S\mathbin{\triangle}\mathcal T$ is one-to-one, which also makes
$C$ symmetric.

Using the symmetry of $C$, Eqs.
\eqref{eq:SM_C_diagonal}--\eqref{eq:SM_C_down} can also be written as
\begin{equation}
 \boxed{
 \begin{aligned}
 C_{\ell,\ell}
 &=\frac{
 \binom{2N-4}{\ell}
 +\binom{2N-4}{\ell-4}
 +6\binom{2N-4}{\ell-2}}
 {\binom{2N}{\ell}},\\[1mm]
 C_{\ell,\ell+2}
 &=-\frac{4\binom{2N-4}{\ell-1}}
 {\sqrt{\binom{2N}{\ell}\binom{2N}{\ell+2}}},
 \qquad
 C_{\ell,\ell-2}
 =-\frac{4\binom{2N-4}{\ell-3}}
 {\sqrt{\binom{2N}{\ell}\binom{2N}{\ell-2}}}.
 \end{aligned}}
 \label{eq:SM_C_main_text}
\end{equation}
These entries determine $\Gamma_{r,m}=C_{2r,2m}$ in Eq.~(7) of the main text.  As in the preceding section,
$\binom ab=0$ when $b<0$ or $b>a$, so the expressions also apply at
the shell boundaries.

\section{Derivation of Eq.~(8) of the main text}
\label{sec:SM_GUE_fidelity_derivation}

In this section, we derive Eq.~(8) for encoding-error-decoding (EED)
dynamics generated by a Gaussian-unitary-ensemble (GUE) Hamiltonian.
We use the notation of the main text and of the preceding derivation of
Eqs.~(4)--(6).  The system contains \(N\) qubits, the logical subsystem
\(X\) contains \(k\) qubits, and the ancilla subsystem
\(\bar X\) contains \(N-k\) qubits.  The Hilbert-space dimension is \(d=2^N\).

\subsection{Annealed fidelity as a ratio of two GUE averages}

The initial state for the GUE Hamiltonian dynamics is the
computational-basis N\'eel state
\begin{equation}
 \ket{\psi_{\mathrm{in}}}
 =\ket{\psi_X^{\mathrm{in}}}
 \otimes\ket{\psi_{\bar X}^{\mathrm{in}}},
 \qquad
 \rho_{\mathrm{in}}
 =\ket{\psi_{\mathrm{in}}}\!\bra{\psi_{\mathrm{in}}}.
 \label{eq:SM_GUE_initial_state}
\end{equation}
Here $\rho_{\mathrm{in}}$ denotes the N\'eel input of this section,
rather than the vacuum input used in the matchgate calculation,
as in the main text.  The ancilla postselection projector is
\begin{equation}
 \rho_{\bar X}^{\mathrm{in}}
 =\ket{\psi_{\bar X}^{\mathrm{in}}}
 \!\bra{\psi_{\bar X}^{\mathrm{in}}},
 \qquad
 \Pi_{\bar X}=\id_X\otimes\rho_{\bar X}^{\mathrm{in}}.
 \label{eq:SM_GUE_QX}
\end{equation}
The error operator is exactly the one used in the preceding section,
\begin{equation}
 \hat K
 =\exp\!\left(-\frac{i\alpha}{2}
 \sum_{j=1}^{N}\sigma_j^z\right).
 \label{eq:SM_GUE_K}
\end{equation}
For
\begin{equation}
 U(t)=e^{-iHt},
 \qquad H\in\mathrm{GUE}(d),
 \label{eq:SM_GUE_U}
\end{equation}
the state after encoding, the error, and decoding is
\begin{equation}
 \ket{\psi_{\mathrm{out}}}=U^\dagger\hat K U\ket{\psi_{\mathrm{in}}}.
 \label{eq:SM_GUE_final_state}
\end{equation}

For a fixed realization of \(H\), the unnormalized fidelity numerator
and the postselection probability are
\begin{align}
 n(H)
 &=\Tr\!\left(
 \rho_{\mathrm{in}}U^\dagger\hat K U
 \rho_{\mathrm{in}}U^\dagger\hat K^\dagger U
 \right),
 \label{eq:SM_GUE_fixed_numerator}\\
 p(H)
 &=\Tr\!\left(
 \Pi_{\bar X}U^\dagger\hat K U
 \rho_{\mathrm{in}}U^\dagger\hat K^\dagger U
 \right).
 \label{eq:SM_GUE_fixed_denominator}
\end{align}
Consequently, the annealed fidelity in Eq.~(8) is
\begin{equation}
 \tilde F_{\mathrm{GUE}}
 =\frac{\mathbb E_H[n(H)]}{\mathbb E_H[p(H)]}.
 \label{eq:SM_GUE_annealed_fidelity}
\end{equation}
As in Eq.~(2), this ratio of separately averaged quantities is
distinct at finite \(N\) from
\(\mathbb E_H[n(H)/p(H)]\).

Both averages are obtained from the common trace
\begin{equation}
 I_B(t,\alpha)
 =\mathbb E_H\Tr\!\left(
 BU^\dagger\hat K U
 \rho_{\mathrm{in}}U^\dagger\hat K^\dagger U
 \right),
 \label{eq:SM_GUE_common_trace}
\end{equation}
where \(B=\rho_{\mathrm{in}}\) gives the numerator and
\(B=\Pi_{\bar X}\) gives the
denominator.

\subsection{GUE eigenbasis average and the Weingarten matrix}

A GUE Hamiltonian can be diagonalized as
\begin{equation}
 H=V\Lambda V^\dagger,
 \qquad
 \Lambda=\operatorname{diag}(\lambda_1,\ldots,\lambda_d),
 \label{eq:SM_GUE_eigendecomposition}
\end{equation}
where \(V\) is Haar distributed over \(U(d)\) and statistically
independent of \(\Lambda\)~\cite{Cotler2017}.  Defining
\begin{equation}
 D(t)=e^{-i\Lambda t},
 \qquad
 U(t)=VD(t)V^\dagger,
 \label{eq:SM_GUE_D}
\end{equation}
we obtain
\begin{align}
 I_B(t,\alpha)
 =\mathbb E_{\Lambda,V}\Tr\big(
 &BVD^\dagger V^\dagger
 \hat KVDV^\dagger
 \rho_{\mathrm{in}}VD^\dagger V^\dagger
 \nonumber\\[-1mm]
 &\times\hat K^\dagger VDV^\dagger
 \big).
 \label{eq:SM_GUE_common_eigenbasis}
\end{align}
Introduce the ordered lists
\begin{equation}
 (A_1,A_2,A_3,A_4)
 =(B,\hat K,\rho_{\mathrm{in}},\hat K^\dagger),
 \qquad
 (\Omega_1,\Omega_2,\Omega_3,\Omega_4)
 =(D^\dagger,D,D^\dagger,D).
 \label{eq:SM_GUE_operator_lists}
\end{equation}

We next fix the permutation convention in Eq.~(8).  For
\(\pi\in S_4\), define
\begin{equation}
 \Tr_{\pi}(O_1,O_2,O_3,O_4)
 =\prod_{c=(i_1\cdots i_s)\in\cyc(\pi)}
 \Tr(O_{i_1}\cdots O_{i_s}).
 \label{eq:SM_GUE_permutation_trace}
\end{equation}
Here \(s\) denotes only the length of the cycle \(c\); it is unrelated
to the Majorana-shell coefficient \(s_{2m}\) used in the derivation of
Eq.~(6).  Let
\begin{equation}
 G_{\sigma,\tau}
 =d^{\#\cyc(\sigma^{-1}\tau)},
 \qquad
 \Wg_d(\sigma,\tau)
 \equiv [G^{-1}]_{\sigma,\tau}
 \label{eq:SM_GUE_Weingarten_matrix}
\end{equation}
be the \(S_4\) Gram and Weingarten matrices.  Equivalently,
\(\Wg_d(\sigma,\tau)=\Wg_d(\sigma^{-1}\tau)\).  Since the
unitary Weingarten function is a class function, this is also equal
to \(\Wg_d(\tau\sigma^{-1})\).

For completeness, we make the index contraction explicit.  At fixed
\(\Lambda\), the integrand in Eq.~\eqref{eq:SM_GUE_common_eigenbasis}
is
\begin{equation}
 \sum_{\{x_r,a_r,b_r,c_r\}}
 \prod_{r=1}^{4}
 (A_r)_{x_ra_r}V_{a_rb_r}
 (\Omega_r)_{b_rc_r}V^*_{x_{r+1}c_r},
 \qquad x_5=x_1.
 \label{eq:SM_GUE_index_form}
\end{equation}
The fourth Haar moment identifies
\(a_r=x_{\sigma(r)+1}\) and \(b_r=c_{\tau(r)}\).  With
\begin{equation}
 \gamma=(1234),
 \label{eq:SM_GUE_gamma}
\end{equation}
and permutation products acting from right to left, so that
\((\gamma\sigma)(r)=\gamma(\sigma(r))\), the \(x\)-index loops give
\(\Tr_{\gamma\sigma}(A_1,A_2,A_3,A_4)\), whereas the spectral
loops give
\(\Tr_{\tau^{-1}}(\Omega_1,\Omega_2,\Omega_3,\Omega_4)\).
Because \(D\) and \(D^\dagger\) are diagonal and mutually commuting,
\begin{equation}
 \Tr_{\tau^{-1}}(D^\dagger,D,D^\dagger,D)
 =\Tr_{\tau}(D^\dagger,D,D^\dagger,D).
 \label{eq:SM_GUE_tau_inverse}
\end{equation}
The eigenbasis average is therefore
\begin{equation}
 I_B(t,\alpha)
 =\sum_{\sigma,\tau\in S_4}
 \Wg_d(\sigma,\tau)
 \Tr_{\gamma\sigma}(B,\hat K,\rho_{\mathrm{in}},\hat K^\dagger)
 \mathcal R_\tau(t),
 \label{eq:SM_GUE_Weingarten_result}
\end{equation}
where
\begin{equation}
 \boxed{
 \mathcal R_\tau(t)
 =\mathbb E_\Lambda\!\left[
 \Tr_{\tau}(D^\dagger,D,D^\dagger,D)
 \right].}
 \label{eq:SM_GUE_Rtau_definition}
\end{equation}
The appearance of \(S_4\) follows from the four copies each of \(V\)
and \(V^\dagger\) in Eq.~\eqref{eq:SM_GUE_common_eigenbasis}.

\subsection{Permutation-resolved GUE spectral factors}

All time dependence is contained in \(\mathcal R_\tau(t)\).  Define
\begin{equation}
 Z_a(t)=\Tr e^{ia\Lambda t},
 \qquad
 Z_0(t)=d,
 \qquad
 Z_{-a}(t)=Z_a^*(t),
 \qquad a\in\mathbb Z.
 \label{eq:SM_GUE_Za}
\end{equation}
The four spectral operators in
Eq.~\eqref{eq:SM_GUE_operator_lists} carry signs
\begin{equation}
 (\epsilon_1,\epsilon_2,\epsilon_3,\epsilon_4)
 =(+1,-1,+1,-1).
 \label{eq:SM_GUE_spectral_signs}
\end{equation}
For a cycle \(c\in\cyc(\tau)\), define the spectral charge
\begin{equation}
 \chi(c)=\sum_{i\in c}\epsilon_i.
 \label{eq:SM_GUE_spectral_charge}
\end{equation}
Then
\begin{equation}
 \boxed{
 \mathcal R_\tau(t)
 =\mathbb E_\Lambda\!\left[
 \prod_{c\in\cyc(\tau)}Z_{\chi(c)}(t)
 \right].}
 \label{eq:SM_GUE_Rtau_charge_form}
\end{equation}

In terms of the standard GUE spectral form factors
\begin{align}
 R_2(t)&=\mathbb E_\Lambda[Z_1(t)Z_{-1}(t)],
 \nonumber\\
 R_4(t)&=\mathbb E_\Lambda[Z_1^2(t)Z_{-1}^2(t)],
 \nonumber\\
 R_{4,1}(t)&=\mathbb E_\Lambda[Z_1^2(t)Z_{-2}(t)],
 \nonumber\\
 R_{4,2}(t)&=\mathbb E_\Lambda[Z_2(t)Z_{-2}(t)]
 =R_2(2t),
 \label{eq:SM_GUE_standard_SFFs}
\end{align}
the 24 spectral permutations reduce to Table~\ref{tab:SM_GUE_Rtau}.
The GUE eigenvalue measure is invariant under
\(\Lambda\mapsto-\Lambda\), so the ensemble-averaged
\(R_{4,1}(t)\) is real.

\begin{table}[htbp]
\caption{Permutation-resolved spectral factors in the convention of
Eq.~\eqref{eq:SM_GUE_Rtau_definition}.}
\label{tab:SM_GUE_Rtau}
\begin{tabular}{lc}
\toprule
Permutation \(\tau\) & \(\mathcal R_\tau(t)\)\\
\midrule
\(e\) & \(R_4(t)\)\\
\((12),(14),(23),(34)\) & \(dR_2(t)\)\\
\((13)\) & \(R_{4,1}^*(t)\)\\
\((24)\) & \(R_{4,1}(t)\)\\
\((12)(34),(14)(23)\) & \(d^2\)\\
\((13)(24)\) & \(R_{4,2}(t)=R_2(2t)\)\\
any 3-cycle & \(R_2(t)\)\\
any 4-cycle & \(d\)\\
\bottomrule
\end{tabular}
\end{table}

Equation~\eqref{eq:SM_GUE_Rtau_charge_form} is exact at finite \(d\).
If one substitutes the closed spectral-kernel expressions of
Ref.~\cite{Cotler2017}, their stated large-\(d\) approximation and
GUE normalization must be used.  A rescaling of \(H\) changes only
the time argument of these spectral factors.

\subsection{Fixed-operator cycle factors for the N\'eel state}

For each site, write
\begin{equation}
 \sigma_j^z\ket{\psi_{\mathrm{in}}}
 =z_j\ket{\psi_{\mathrm{in}}},
 \qquad z_j\in\{+1,-1\},
 \label{eq:SM_GUE_zj}
\end{equation}
We define the total and ancilla magnetizations of the selected
N\'eel configuration as
\begin{equation}
 \mathcal M_0=\sum_{j=1}^{N}z_j,
 \qquad
 \mathcal M_{\bar X}=\sum_{j\in\bar X}z_j.
 \label{eq:SM_GUE_magnetizations}
\end{equation}
The calligraphic symbol \(\mathcal M\) is used here to avoid any
confusion with the matchgate subscript \(M\) in \(U_{\mathrm M}\) and
\(\Phi_{\mathrm M}^{(2)}\) in the preceding section.

For a cycle \(c\in\cyc(\gamma\sigma)\), define the net error charge
\begin{equation}
 q(c)=\ind_{\{2\in c\}}-\ind_{\{4\in c\}}
 \in\{-1,0,+1\}.
 \label{eq:SM_GUE_error_charge}
\end{equation}
The labels 2 and 4 refer to \(\hat K\) and
\(\hat K^\dagger=\hat K^{-1}\), respectively, in the ordered list
in Eq.~\eqref{eq:SM_GUE_operator_lists}.  All fixed operators are
diagonal in the computational basis, so the error operators in a
cycle reduce to \(\hat K^{q(c)}\).

Only three elementary traces are required:
\begin{align}
 \eta_q
 &\equiv
 \bra{\psi_{\mathrm{in}}}\hat K^q
 \ket{\psi_{\mathrm{in}}}
 =e^{-iq\alpha\mathcal M_0/2},
 \label{eq:SM_GUE_eta}\\
 \kappa_q
 &\equiv\Tr(\hat K^q)
 =\left[2\cos\!\left(\frac{q\alpha}{2}\right)\right]^N
 =d\cos^N\!\left(\frac{q\alpha}{2}\right),
 \label{eq:SM_GUE_kappa}\\
 \mu_q
 &\equiv\Tr(\Pi_{\bar X}\hat K^q)
 \nonumber\\
 &=\Tr_X(\hat K_X^q)
 \bra{\psi_{\bar X}^{\mathrm{in}}}\hat K_{\bar X}^q
 \ket{\psi_{\bar X}^{\mathrm{in}}}
 \nonumber\\
 &=2^{k}\cos^{k}\!\left(\frac{q\alpha}{2}\right)
 e^{-iq\alpha\mathcal M_{\bar X}/2}.
 \label{eq:SM_GUE_mu}
\end{align}
In particular,
\(\eta_0=1\), \(\kappa_0=d\), and \(\mu_0=2^{k}\).

For the numerator, \(B=\rho_{\mathrm{in}}\), so indices 1 and 3 in
Eq.~\eqref{eq:SM_GUE_operator_lists} both represent
\(\rho_{\mathrm{in}}\).
Since
\begin{equation}
 (\rho_{\mathrm{in}})^2=\rho_{\mathrm{in}},
 \qquad
 [\rho_{\mathrm{in}},\hat K]=0,
 \label{eq:SM_GUE_rho_properties}
\end{equation}
every cycle containing index 1 or 3 gives \(\eta_{q(c)}\), whereas
a cycle containing neither gives \(\kappa_{q(c)}\).  Hence the
numerator cycle factor in Eq.~(8) is
\begin{equation}
 \boxed{
 g_{\mathrm N}(c)=
 \begin{cases}
 \eta_{q(c)},&c\cap\{1,3\}\neq\varnothing,\\[1mm]
 \kappa_{q(c)},&c\cap\{1,3\}=\varnothing.
 \end{cases}}
 \label{eq:SM_GUE_gN}
\end{equation}
It follows that
\begin{equation}
 \mathbb E_H[n(H)]
 =\sum_{\sigma,\tau\in S_4}
 \Wg_d(\sigma,\tau)
 \left[\prod_{c\in\cyc(\gamma\sigma)}
 g_{\mathrm N}(c)\right]\mathcal R_\tau(t).
 \label{eq:SM_GUE_final_numerator}
\end{equation}

For the denominator, \(B=\Pi_{\bar X}\).  The identities
\begin{equation}
 \Pi_{\bar X}\rho_{\mathrm{in}}
 =\rho_{\mathrm{in}}\Pi_{\bar X}
 =\rho_{\mathrm{in}}
 \label{eq:SM_GUE_QX_rho}
\end{equation}
give three mutually exclusive cases.  A cycle containing index 3
contains \(\rho_{\mathrm{in}}\) and gives \(\eta_{q(c)}\), irrespective of
whether it also contains index 1.  A cycle containing index 1 but
not index 3 contains \(\Pi_{\bar X}\) without
\(\rho_{\mathrm{in}}\) and gives
\(\mu_{q(c)}\).  A cycle containing neither index gives
\(\kappa_{q(c)}\).  Therefore,
\begin{equation}
 \boxed{
 g_{\mathrm D}(c)=
 \begin{cases}
 \eta_{q(c)},&3\in c,\\[1mm]
 \mu_{q(c)},&1\in c\ \text{and}\ 3\notin c,\\[1mm]
 \kappa_{q(c)},&1\notin c\ \text{and}\ 3\notin c.
 \end{cases}}
 \label{eq:SM_GUE_gD}
\end{equation}
and
\begin{equation}
 \mathbb E_H[p(H)]
 =\sum_{\sigma,\tau\in S_4}
 \Wg_d(\sigma,\tau)
 \left[\prod_{c\in\cyc(\gamma\sigma)}
 g_{\mathrm D}(c)\right]\mathcal R_\tau(t).
 \label{eq:SM_GUE_final_denominator}
\end{equation}

Combining Eqs.~\eqref{eq:SM_GUE_final_numerator} and
\eqref{eq:SM_GUE_final_denominator} gives Eq.~(8) of the main text:
\begin{equation}
 \boxed{
 \tilde F_{\mathrm{GUE}}
 =
 \frac{\displaystyle
 \sum_{\sigma,\tau\in S_4}
 \Wg_d(\sigma,\tau)
 \left[\prod_{c\in\cyc(\gamma\sigma)}
 g_{\mathrm N}(c)\right]\mathcal R_\tau(t)}
 {\displaystyle
 \sum_{\sigma,\tau\in S_4}
 \Wg_d(\sigma,\tau)
 \left[\prod_{c\in\cyc(\gamma\sigma)}
 g_{\mathrm D}(c)\right]\mathcal R_\tau(t)}.}
 \label{eq:SM_GUE_final_fidelity}
\end{equation}
The lowercase cycle factors \(g_{\mathrm N}\) and \(g_{\mathrm D}\)
in Eq.~\eqref{eq:SM_GUE_final_fidelity} are distinct from the
Majorana-shell boundary coefficients \(N_{2m}\) and
\(D_{2m}\) in Eq.~(6).

For the balanced N\'eel state used in the main text,
\(\mathcal M_0=0\), and hence
\begin{equation}
 \eta_q=1.
 \label{eq:SM_GUE_balanced_eta}
\end{equation}
If \(X\) and \(\bar X\) each contain an even number of consecutive
sites of the alternating N\'eel pattern, as for the half-system
bipartitions used in the calculation, then
\(\mathcal M_{\bar X}=0\) as well and
\begin{equation}
 \mu_q
 =2^{k}\cos^{k}\!\left(\frac{q\alpha}{2}\right).
 \label{eq:SM_GUE_balanced_mu}
\end{equation}
Keeping \(\mathcal M_{\bar X}\) in Eq.~\eqref{eq:SM_GUE_mu} makes
the result valid for a generic placement or parity of \(X\).

\section{Numerical implementation of random matchgate circuits and self-averaging of the EED fidelity}
\label{sec:SM_matchgate_numerics}

This section describes the numerical implementation used to sample the
random matchgate encoder and to evaluate the sample-resolved EED fidelity.
We first specify the Hurwitz sampling of Haar-random active-FLO circuits,
their implementation in terms of elementary qubit gates, and the exact
state-vector evaluation of the numerator, postselection probability, and
fidelity defined in Sec.~\ref{sec:matchgate_fidelity_derivation}.  The
resulting realization-resolved data provide the input for the
self-averaging analysis described in the second part of this section.

\subsection{Exact state-vector simulation of Haar-random matchgate circuits}
\label{subsec:SM_random_matchgate_numerics}

\subsubsection{Majorana convention and elementary rotations}

We use the Jordan--Wigner Majorana operators
\begin{equation}
 \gamma_{2j-1}
 =\left(\prod_{\ell<j}\sigma_\ell^z\right)\sigma_j^x,
 \qquad
 \gamma_{2j}
 =\left(\prod_{\ell<j}\sigma_\ell^z\right)\sigma_j^y,
 \qquad j=1,\ldots,N,
 \label{eq:SM_numerics_JW}
\end{equation}
which obey
\(\sigma_j^z=-i\gamma_{2j-1}\gamma_{2j}\), in agreement with
Eq.~\eqref{eq:SM_majorana_basis}.  The random active-FLO circuit is built
from adjacent-Majorana rotations
\begin{equation}
 U_{R_j}(\theta)
 =\exp\!\left(\frac{\theta}{2}\gamma_j\gamma_{j+1}\right),
 \qquad j=1,\ldots,2N-1.
 \label{eq:SM_numerics_adjacent_rotation}
\end{equation}
In the qubit representation of Eq.~\eqref{eq:SM_numerics_JW}, these are
implemented as
\begin{equation}
 U_{R_j}(\theta)=
 \begin{cases}
 \exp\!\left(+\dfrac{i\theta}{2}\sigma_q^z\right),
 &j=2q-1,\\[2mm]
 \exp\!\left(+\dfrac{i\theta}{2}
 \sigma_q^x\sigma_{q+1}^x\right),
 &j=2q.
 \end{cases}
 \label{eq:SM_numerics_qubit_rotations}
\end{equation}
Thus only single-qubit \(Z\) rotations and nearest-neighbor \(XX\)
rotations are required.  The signs and half-angle convention in
Eq.~\eqref{eq:SM_numerics_qubit_rotations} are fixed by
\(\gamma_j\gamma_{j+1}=i\sigma_q^z\) for odd \(j\) and
\(\gamma_j\gamma_{j+1}=i\sigma_q^x\sigma_{q+1}^x\) for even \(j\).

\subsubsection{Hurwitz sampling of Haar-random active-FLO circuits}

To generate Haar-random active fermionic linear optics (FLO) unitaries, we employ the triangular Hurwitz parameterization described in Ref.~\cite{Braccia2025}.  For every
\(\ell=2,\ldots,2N\), the circuit contains the \(\ell-1\) angles
\(\theta_{r,\ell}\), with \(r=1,\ldots,\ell-1\).  The first angle in each
column is uniform,
\begin{equation}
 \theta_{1,\ell}\sim\operatorname{Uniform}[0,2\pi),
 \label{eq:SM_numerics_uniform_angle}
\end{equation}
whereas, for \(r=2,\ldots,\ell-1\),
\begin{equation}
 p_r(\theta)
 =\frac{\Gamma[(r+1)/2]}
 {\sqrt{\pi}\,\Gamma(r/2)}\sin^{r-1}\!\theta,
 \qquad 0\leq\theta\leq\pi.
 \label{eq:SM_numerics_sine_density}
\end{equation}
In the implementation, Eq.~\eqref{eq:SM_numerics_sine_density} is sampled
without rejection by drawing
\begin{equation}
 Y\sim\operatorname{Beta}\!\left(\frac r2,\frac r2\right),
 \qquad
 \theta_{r,\ell}=\arccos(2Y-1).
 \label{eq:SM_numerics_beta_sampling}
\end{equation}
The total number of independently sampled continuous parameters is
\begin{equation}
 \sum_{\ell=2}^{2N}(\ell-1)=N(2N-1).
 \label{eq:SM_numerics_parameter_count}
\end{equation}

To state the gate order without ambiguity from matrix-product
conventions, the state-vector routine performs the following nested
updates:
\begin{equation}
 \text{for }j=1,\ldots,2N-1,\quad
 \ell=j+1,\quad\text{and }r=j,j-1,\ldots,1:
 \qquad
 \ket{\psi}\leftarrow
 U_{R_r}(\theta_{r,\ell})\ket{\psi}.
 \label{eq:SM_numerics_forward_loop}
\end{equation}
This gate order is an equivalent Hurwitz word and generates the desired
Haar-random active-FLO ensemble.  The inverse circuit is applied by reversing both
loops and changing the angle signs,
\begin{equation}
 \text{for }j=2N-1,\ldots,1,\quad
 \ell=j+1,\quad\text{and }r=1,2,\ldots,j:
 \qquad
 \ket{\psi}\leftarrow
 U_{R_r}(-\theta_{r,\ell})\ket{\psi}.
 \label{eq:SM_numerics_inverse_loop}
\end{equation}
Equations~\eqref{eq:SM_numerics_forward_loop} and
\eqref{eq:SM_numerics_inverse_loop} therefore implement exactly
\(U_{\mathrm{FLO}}\) and \(U_{\mathrm{FLO}}^\dagger\) for every
stored set of angles.  Although the triangular circuit has
\(O(N^2)\) gates and nonoptimal depth, its output ensemble is identical
to that of the depth-optimized construction of
Ref.~\cite{Braccia2025}.

\subsubsection{EED evolution and realization-resolved observables}

For a single realization \(a\), we begin with the same state and error operator as
in Sec.~\ref{sec:matchgate_fidelity_derivation},
\begin{equation}
 \ket{\psi_{\mathrm{in}}}=\ket{0}^{\otimes N},
 \qquad
 \hat K=\prod_{j=1}^{N}
 \exp\!\left(-\frac{i\alpha}{2}\sigma_j^z\right),
 \label{eq:SM_numerics_initial_error}
\end{equation}
and apply
\begin{equation}
 \ket{\psi_{\mathrm{out}}^{(a)}}
 =U_{\mathrm M,a}^\dagger\hat K U_{\mathrm M,a}\ket{0}^{\otimes N},
 \qquad U_{\mathrm M,a}=\gamma_1^{\xi_a}U_{\mathrm{FLO},a},
 \label{eq:SM_numerics_final_state}
\end{equation}
Here $\xi_a\in\{0,1\}$ is sampled independently with equal
probabilities.  The fixed improper Majorana reflection
$\gamma_1=X_1$ extends the Haar-random $SO(2N)$ active-FLO draws
to the full $O(2N)$ Gaussian ensemble used in the main text.
Thus the state-vector routines apply
$U_{\mathrm{FLO},a}$ and $\gamma_1^{\xi_a}$, followed by $\hat K$,
and then $\gamma_1^{\xi_a}$ and the stored inverse
$U_{\mathrm{FLO},a}^\dagger$.  For the undoped
matchgate calculation there is a single Haar-random Gaussian encoder,
and no non-Gaussian gates are applied.

Using \(\Pi_{\bar X}=\id_X\otimes\rho_{\bar X}^{\mathrm{in}}\) from
Eq.~\eqref{eq:SM_initial_data}, we calculate the following quantities: 
\begin{align}
 n_a
 &=\left|\braket{0^{\otimes N}|\psi_{\mathrm{out}}^{(a)}}\right|^2,
 \label{eq:SM_numerics_num}\\
 p_a
 &=\bra{\psi_{\mathrm{out}}^{(a)}}\Pi_{\bar X}\ket{\psi_{\mathrm{out}}^{(a)}}
 =\sum_{x_X\in\{0,1\}^{k}}
 \left|\braket{x_X,0_{\bar X}|\psi_{\mathrm{out}}^{(a)}}\right|^2,
 \label{eq:SM_numerics_den}\\
 F_a&=\frac{n_a}{p_a}.
 \label{eq:SM_numerics_fidelity}
\end{align}
These are precisely the realization-resolved quantities
\(n(U_{\mathrm M})\), \(p(U_{\mathrm M})\), and \(F(U_{\mathrm M})\) defined in
Eqs.~\eqref{eq:SM_fixed_numerator} and
\eqref{eq:SM_fixed_denominator}. We consider $10^{4}$ circuit realizations in the calculation. 

\begin{figure}[]
	\centering
	\includegraphics[width=0.5\linewidth]{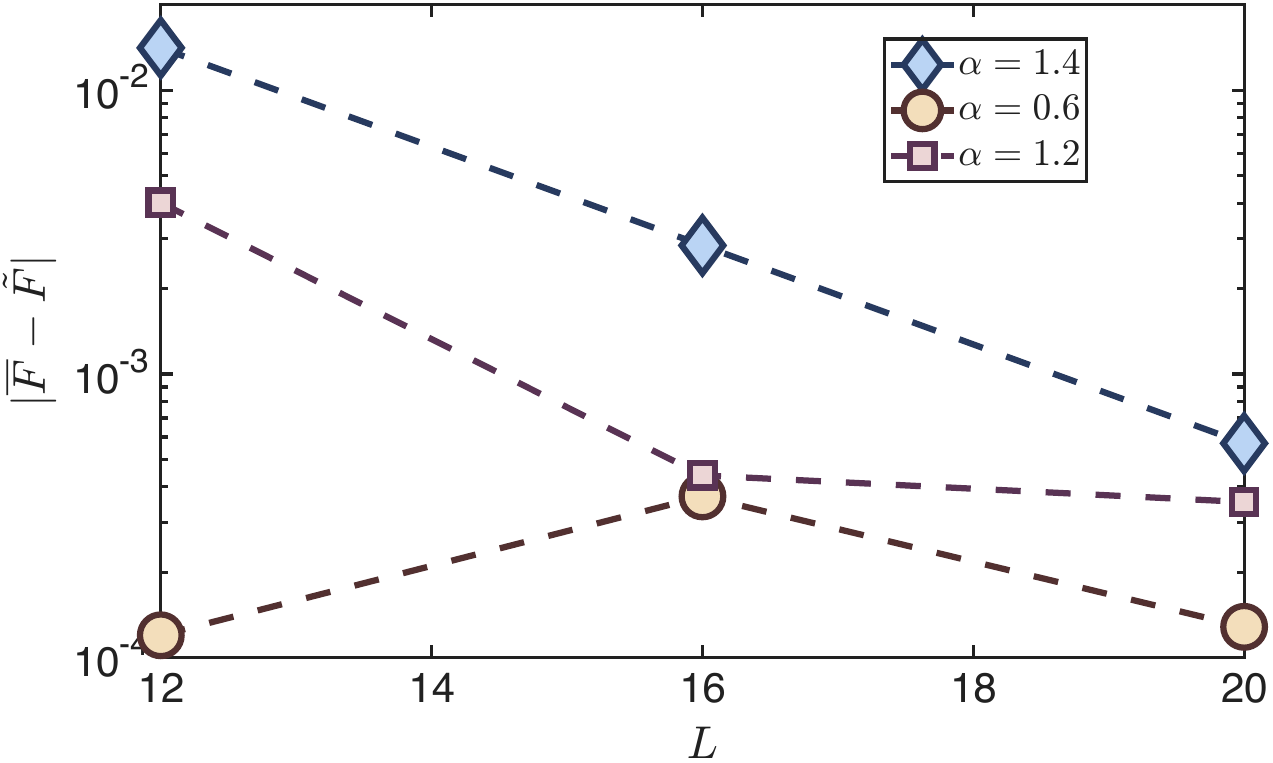}\\
	\caption{For EED matchgate circuits without non-Gaussian doping gates, the difference between the quenched and annealed fidelity $|\overline{F}-\tilde F_{\mathrm M}|$ as a function of system size $N$ with different error strengths $\alpha$. The horizontal-axis label $L$ denotes $N$. }\label{fig_s1}
\end{figure}

\subsection{Self-averaging of the sample-resolved fidelity}
\label{subsec:SM_matchgate_self_averaging}

Here, we study the difference between the quenched fidelity and annealed fidelity, which are defined by 
\begin{align}
\overline{F} = \mathbb{E}_{a}[\frac{n_{a}}{p_{a}}],
\end{align}
and
\begin{align}
\tilde F_{\mathrm M} = \frac{\mathbb{E}_{a}[n_{a}]}{\mathbb{E}_{a}[p_{a}]},
\end{align}
respectively. We show the results of the fidelity difference $|\overline{F}-\tilde F_{\mathrm M}|$ for the EED matchgate circuits without non-Gaussian doping gates in Fig.~\ref{fig_s1}. At $N=20$, this difference is below $10^{-3}$ for all error strengths shown, supporting the annealed approximation to the quenched fidelity in this regime.